\documentclass{article}
 \pdfoutput=1
\usepackage{amsmath, amssymb, amsthm, amscd, amsfonts}
\usepackage{graphicx}
\usepackage{enumitem}
\usepackage{times}
\usepackage{xspace}
\usepackage{bbm}
\usepackage{thmtools,thm-restate}
\usepackage{import}
\usepackage{multicol}

\usepackage{algorithm, algorithmicx, algcompatible}
\usepackage[noend]{algpseudocode}

\theoremstyle{plain} \numberwithin{equation}{section}
\newtheorem{theorem}{Theorem}[section]
\newtheorem{observation}{Observation}

\theoremstyle{definition}
\newtheorem{definition}[theorem]{Definition}

\newtheorem{remark}[theorem]{Remark}

\newtheorem{type}[theorem]{Object Type}

\newcommand{\true}[0]{\mbox{\em true}} 
\newcommand{\false}[0]{\mbox{\em false}} 

\newcommand{\N}[0]{\mathbb{N}} 

\newcommand{\nullset}[0]{\varnothing}

\newcommand{\cmdprocedure}[0]{\textbf{procedure}\xspace}
\newcommand{\cmdprogram}[0]{\textbf{program}\xspace}

\newcommand{\cmdfor}[0]{\textbf{for}\xspace}
\newcommand{\cmdwhile}[0]{\textbf{while}\xspace}
\newcommand{\cmddo}[0]{\textbf{do}\xspace}

\newcommand{\cmdif}[0]{\textbf{if}\xspace}
\newcommand{\cmdthen}[0]{\textbf{then}\xspace}
\newcommand{\cmdelse}[0]{\textbf{else}\xspace}
\newcommand{\cmdelif}[0]{\textbf{else if}\xspace}

\newcommand{\cmdgoto}[0]{\textbf{goto}\xspace}
\newcommand{\cmdreturn}[0]{\textbf{return}\xspace}

\newcounter{int}
\newcommand{\tab}[1][1]{\setcounter{int}{0}\loop\hspace{\algorithmicindent}\addtocounter{int}{1}\ifnum\value{int}<#1\repeat}

\newcommand{\ack}[0]{\mbox{\em ack}} 

\newcommand{\CAS}[1]{\Call{Cas}{#1}}

\newcommand{\FAInc}[1]{\Call{F\&Inc}{#1}}

\newcommand{\FAS}[1]{\Call{Fas}{#1}}

\newcommand{\Find}[0]{\textsc{Find}} 
\newcommand{\Unite}[0]{\textsc{Unite}}

\newcommand{\Enqueue}[0]{\textsc{Enqueue}} 
\newcommand{\Dequeue}[0]{\textsc{Dequeue}} 

\newcommand{\Write}[0]{\textsc{Write}} 
\newcommand{\Scan}[0]{\textsc{Scan}}

\newcommand{\Ocal}[0]{\mathcal{O}}
\newcommand{\A}[0]{\mathcal{A}}

\newcommand{\C}[0]{\mathcal{C}} 
\newcommand{\M}[0]{\mathcal{M}} 
 
\newcommand{\Rcal}[0]{\mathcal{R}}
\newcommand{\Lcal}[0]{\mathcal{L}}

\newcommand{\ol}[0]{\overline}

\newcommand{\Mfull}[0]{\M^*} 
\newcommand{\Ofull}[0]{\Ocal^*} 
\newcommand{\Rfull}[0]{R^*} 
\newcommand{\Cfull}[0]{C^*} 
\newcommand{\ellfull}[0]{\ell^*}

\newcommand{\Mpart}[0]{\mathcal{M}'}
\newcommand{\Opart}[0]{\Ocal'} 
\newcommand{\Rpart}[0]{R'} 
\newcommand{\Cpart}[0]{C'} 
\newcommand{\ellpart}[0]{\ell'}

\newcommand{\Omegaaug}[0]{\ol{\Omega}} 
\newcommand{\Oaug}[0]{\ol{\Ocal}} 
\newcommand{\Raug}[0]{\ol{R}} 
\newcommand{\Caug}[0]{\ol{C}} 
\newcommand{\ellaug}[0]{\ol{\ell}}

\newcommand{\augColor}{purple}
\newcommand{\augment}[1]{{\color{\augColor}#1}}

\newcommand{\I}[0]{\mathcal{I}}

\usepackage{xpatch}
\makeatletter
\xpatchcmd\ALG@step{\arabic{ALG@line}}{\fmtlinenumber{ALG@line}}{}{}
\makeatother 
\let\fmtlinenumber\arabic 
\newcommand\mathalph[1]{$\alph{#1}$} 

\usepackage{url}
\usepackage{xcolor}
\usepackage{graphicx}
\usepackage[margin=1in]{geometry}

\title{A Universal Technique for Machine-Certified Proofs of Linearizable Algorithms\footnote{This paper contains material presented in Siddhartha Jayanti's Ph.D. thesis \cite{SiddharthaPhDThesis}, Ugur Yavuz's bachelor's and master's theses \cite{UgurUndergraduateThesis, UgurMastersThesis}, and Lizzie Hernandez's bachelor's thesis \cite{LizzieUndergraduateThesis}.}} 

\author{
Prasad Jayanti\footnote{Dartmouth College; prasad.jayanti@dartmouth.edu} 
\and 
Siddhartha V. Jayanti\footnote{Google Research and MIT; sjayanti@google.com, siddhartha@csail.mit.edu} 
\and 
Ugur Y. Yavuz\footnote{Boston University; uyyavuz@bu.edu} 
\and 
Lizzie Hernandez\footnote{Microsoft; lizzie.hernandez.videa.22@dartmouth.edu}
}

\date{January 10, 2023}

\begin{document}

\maketitle
\begin{abstract}
    Linearizability has been the long standing gold standard for consistency in concurrent data structures.
    However, proofs of linearizability can be long and intricate, hard to produce, and extremely time consuming even to verify.
    In this work, we address this issue by introducing simple {\em universal}, {\em sound}, and {\em complete} proof methods for producing machine-verifiable proofs of linearizability and its close cousin, strong linearizability.
    Universality means that our method works for any object type; 
    soundness means that an algorithm can be proved correct by our method only if it is linearizable (resp. strong linearizable); and 
    completeness means that any linearizable (resp. strong linearizable) implementation can be proved so using our method.
    We demonstrate the simplicity and power of our method by producing proofs of linearizability for the Herlihy-Wing queue and Jayanti's single-scanner snapshot, as well as a proof of strong linearizability of the Jayanti-Tarjan union-find object.
    All three of these proofs are machine-verified by TLAPS (the Temporal Logic of Actions Proof System).
\end{abstract}

\section{Introduction}

Data structures that organize, store, and quickly recall important pieces of information are the fundamental building blocks behind fast algorithms.
Thus, efficient and rigorously proved data structures are fundamental to reliable algorithm design.
The task of designing such data structures for shared-memory multiprocessors, however, is notoriously difficult.
Due to asynchrony, a $t$ step algorithm for even just two processes has $2^t$, i.e. exponentially many, possible executions depending on how the steps of the processes interleave.
In fact, even deterministic concurrent algorithms have uncountably many possible infinite executions, as opposed to the single possible execution of a deterministic sequential algorithm.
Designing algorithms that are correct in all of these executions is a grueling task, and thus, even mission-critical concurrent code often suffers from elusive races.
For example, a subtle priority inversion bug in its concurrent code crashed the Pathfinder Rover days after its deployment on Mars and jeopardized the entire multi-million dollar NASA space mission \cite{MarsRover}.
Examples of errors in published concurrent data structures are also not left wanting \cite{Mistake1, Mistake2}.

\subsection{Understanding the Problem}

Rigorously proving the correctness of a data structure implementation $\Ocal$ consists of two steps:
\begin{enumerate}
    \item 
    A {\em prover}, often the algorithm designer, deeply studies the implementation $\Ocal$ and produces a {\em proposed proof} $P$.
    \item
    A {\em verifier} evaluates the proposed proof $P$ by confirming that each claim made in the proof is mathematically justified, and that the resultant chain of reasoning constitutes a legitimate proof of $\Ocal$'s correctness.
\end{enumerate}
In general, proving any type of correctness of any algorithm can be inherently intellectually difficult and time consuming, since it requires the prover to contemplate the algorithm deeply, understand why it is correct, and express this understanding in the methodical language of mathematics.
Verifying many types of proofs can be rather easier, since it is in its essence a mechanical check.

In this work, we focus on concurrent data structures for asynchronous shared-memory multiprocessors.
In particular, {\em linearizability} \cite{HW90}, which states that data structure operations must appear to take place atomically even in the face of tremendous concurrency, has been the longstanding gold standard for concurrent data structure correctness.
Its close cousin {\em strong linearizability} \cite{StrongLinearizability}, which ensures that even the hyper-properties of the data structure match those of an atomic object has also garnered a lot of recent interest \cite{Attiya19}.

When the correctness condition being proved is linearizability, we observe, in practice, that even the verification step can be taxing and time-consuming.
Firstly, the proposed proof $P$ can often run for several tens of pages of a dense research paper.
For example, the original paper on linearizability \cite{HW90} contains a short seven line implementation of a queue.
To prove its linearizability, the authors propose an eighteen line invariant.
The proof of correctness of this invariant and its entailment of the queue's linearizability are presented in a 14 page technical report.
Those familiar with this queue implementation know that it, like many linearizable algorithms, is inherently ``very tricky'' to prove correct, and thus it is commendable that the authors, Herlihy and Wing, were able to design this algorithm and give a proof of it.
Nevertheless, the subtlety and sheer length of the proof $P$, makes the job of a conscientious verifier quite hard.

Depending on the complexity of an algorithm and its proof, the mere process of verifying the mathematical validity of the linearizability proof can take hours, days, or even weeks.
In some cases, due to the difficulty inherent to writing such long and intricate proofs, provers resort to making high-level ``hand-wavy'' arguments, or omit proofs altogether.
This makes the verifier's job more difficult to impossible.
In other cases, the verifier may be a conference reviewer who does not have the time to verify a long proof and thus must skim the details of the proof which may contain errors;
or the end-user of the algorithm may not be a verifier, but an engineer who may not have the time or the mathematical preparation to independently verify the algorithm's correctness before using it in a deployed system.

For all of these reasons, it is not shocking that mistakes in concurrent algorithms are so prevalent, even in mission critical deployed code.
To avoid such critical errors, we propose designing concurrent data structures whose correctness is {\em machine-verified} to limit the scope for human error.

\subsection{Our Work}

Informally, an object $\Ocal$ is linearizable, if {\em for all} finite runs $R$ of any algorithm $\A$ that uses $\Ocal$, and every operation $op$ that is performed on $\Ocal$ in the run $R$, {\em there exists} a point in time between $op$'s invocation and return where it ``appears to take place instantaneously''.
This definition, in particular the non-constructive existential quantifier ``{\em there exists}'' inside the universal quantification, makes it difficult to prove linearizability.
This difficulty is only exacerbated if the proof must be simple enough for a machine to verify.
In fact, if approached na\"{\i}vely, the prover would need to map each run of the algorithm to a linearization, i.e., a description of where in its invocation-response time interval each operation ``appears to take place instantaneously'', and then prove that each such mapping is legitimate.
This is a difficult task, given that it is known that proving even a single fixed run $R$ linearizable is NP-hard \cite{LinearizabilityNPComplete}.

Our goal however, is to devise a method for proving linearizability that not only works for a single implementation, or even a single type, but
to devise a method that is {\em universal} and {\em complete}.
By universal, we mean that our method should be powerful enough to allow for a proof of linearizability for implementations of {\em any} object type.
By complete, we mean that {\em any linearizable implementation}, regardless of how complex its expression or linearization structure, must be provable by our method.
Of course, our method will also be {\em sound}, meaning that any argument that is given using our method is indeed a correct mathematical proof of linearizability.
Finally, we ensure that our method enables machine verifiable proofs by currently available proof assistants, which are generally built to verify proofs of simple program invariants.

\subsubsection{Our Contributions}

\begin{enumerate}
    \item 
    We develop a rigorous {\em universal}, {\em sound}, and {\em complete} method for proving linearizability.
    In particular, we define a universal transformation that takes an arbitrary implementation $\Ocal$ of an arbitrary type $\tau$, and outputs an algorithm $\A^*$, called the {\em tracker}, and a simple invariant $\I^*$, and prove a theorem that:
    
    {
    \centering
    \underline{$\Ocal$ is a linearizable implementation of type $\tau$ {\em if and only if} $\I^*$ is an invariant of $\A^*$.}
    \par
    }
    
    (Thus, we can produce a machine-certified proof that $\I^*$ is an invariant of $\A^*$ to establish that $\Ocal$ is linearizable.)

    \item
    In fact, we give a family of transformations that each output different algorithms $\A'$, called {\em partial trackers}, with associated invariants $\I'$,
    and prove that:
    
    {
    \centering
    \underline{$\Ocal$ is a linearizable {\em iff} some partial tracker $\A'$ has its associated $\I'$ as an invariant.}
    \par
    }

    
    
    

    \item
    We develop a rigorous {\em universal}, {\em sound}, and {\em complete} method for proving strong linearizability.
    In particular, we show that for each partial tracker $\A'$, there is an alternate associated invariant $\I''$, and we prove that:
 
     {\centering
    \underline{$\Ocal$ is strongly linearizable {\em iff} some partial tracker $\A'$ has its associated $\I''$ as an invariant.}
    \par}


    \item
    Finally, we demonstrate the power of our methods by producing machine-certified proofs of linearizability and strong linearizability for some notable data structures.
    In particular, we prove the linearizability of two famous data structures:
    (1) the aforementioned Herlihy-Wing queue \cite{HW90}, which is notorious for being hard to prove correct \cite{BackwardSimulationLinearizability}; and
    (2) Jayanti's single-writer, single-scanner snapshot algorithm, in which processes play asymmetric roles \cite{PJayanti05}.
    We also prove the strong linearizability of the Jayanti-Tarjan union-find object \cite{JT16, JayantiTarjanBoix, JT21}, which is known to be the fastest algorithm for computing connected components on CPUs and GPUs \cite{dhulipala2020connectit, GpuUnionFind}.
    All our proofs have been certified by the proof assistant TLAPS (temporal logic of actions proof system) \cite{TLAPS}, and are publicly available on GitHub\footnote{The proofs are available at: \url{https://github.com/uguryavuz/machine-certified-linearizability}}. 

\end{enumerate}

\section{Related Work}

Herlihy and Wing's landmark paper that introduced linearizability also seeded the discussion on proof techniques for linearizability \cite{HW90}.
In particular, their paper introduced the concept of {\em possibilities}, i.e. the notion that we can imagine several different orders in which partially completed operations could return in the future, and consider linearizations that are consistent with such possibilities.
They expanded on this framework in a related publication \cite{HW87}.
Their initial ideas in these papers saw fruition in their proof of linearizability of the tricky Herlihy-Wing queue implementation.





Most work on machine-assisted reasoning about linearizability can be classified into three groups: model checking, proofs of particular objects, and more general techniques.
Several works model check algorithms \cite{ModelChecking1, ModelChecking2, ModelChecking3}.
Model checking {\em does not} prove correctness; rather, it mechanically simulates small runs of the algorithm and returns a counter-example if it finds one.
Some works machine-verify the linearizability of {\em specific implementations}, such as: Scott's lock-free queue \cite{ScottQueueProof},
a concurrent open addressing hash table \cite{HashTableProof},
and a list-based set algorithm \cite{ConcurrentListProof}.
Various other algorithms have also been proved linearizable using shape analyses that examine the pointer structures within an object \cite{Shape1, Shape2, Shape3}.
Most general purpose techniques for proving linearizability, such as \cite{Vafeiadis10, VafeiadisHerlihyHoareShapiro2006}, are sound but not complete or universal, i.e.,
they are targeted at showing the linearizability of a limited class of algorithms, rather than any linearizable implementation of any type.

To our knowledge, the only previous sound and complete technique is by Schellhorn et al. \cite{BackwardSimulationLinearizability}.
This technique mechanizes Herlihy and Wing's original possibilities proof strategy through the technology of observational refinement mappings and backward simulations.
The authors give a single example demonstration of their technique---a mechanized proof of the Herlihy-Wing queue verified by the proof assistant KIV.
In contrast to Schellhorn et al.'s technique, our method for proving linearizability only requires familiarity with the notion of an invariant.

Dongol and Derrick wrote an extensive survey on machine assisted proofs of (standard) linearizability \cite{LinearizabilityProofsSurvey}.
To the best of our knowledge, we are the first to introduce machine-verifiable proof methods for strong linearizability.

\section{Model and Definitions}
\label{sec:model}

A {\em concurrent system} consists of a set of asynchronous processes, $\Pi$, that communicate through operations on a set of shared objects, $\Omega$.
Each process has a distinct name and a set of private registers, including a {\em program counter}.
Each object has a distinct name and a {\em type}, which specifies the operation's supported by the object and how these operations behave, i.e. how each operation changes the object's state and what response it returns.
An {\em algorithm} specifies a program for each process, and an initial state for each object.
An algorithm's {\em execution} proceeds in steps.
In a {\em step}, any one process atomically executes the line pointed to by its program counter.
It is common for algorithms to restrict each line in a program to apply at most one operation on a shared object; however, to ensure that our results apply to a wider class of algorithms, we do not impose such a restriction.
In an {\em asynchronous} execution of the algorithm, a (possibly) {\em adversarial scheduler} decides which process $\pi \in \Pi$ will execute the next step in its algorithm at each discrete time step. 
We formalize and expound these notions below.

\begin{definition}[object type]
	An {\em object type} $\tau$ consists of the following {\em components}:
	\begin{itemize}
		\item 
		a set of {\em states} $\Sigma$ that the object can be in.
		
		\item
		a set of {\em operations} $OP$ that can be invoked on the object.
		
		\item
		for each $op \in OP$, a set of {\em arguments} $ARG_{op}$ that the operation $op$ can be called with.
		
		\item
		a set of {\em responses} $RES$, a.k.a. {\em return values}.
		
		\item
		a {\em transition function} $\delta(\sigma, \pi, op, arg)$ that outputs the new state $\sigma'$ and the return value $res$ that result when the operation $op$ with argument $arg$ is performed by process $\pi$ while the object is in state $\sigma$.  
		Formally, the transition function is 
		$$\delta: \Sigma \times \Pi \times\{ (op, arg) \mid op \in OP, arg \in ARG_{op}\} \to \Sigma \times RES$$	
	\end{itemize}
\end{definition}

\begin{remark}
	Operations that require ``no argument'' (i.e., $read$) are modeled as taking an argument from a singleton set (i.e., $ARG_{read} = \{\bot\}$).
	Similarly, operations that return ``no result'' (i.e., $write$) are modeled as returning the result $\ack$.
\end{remark}


\noindent
As an example of an object type, we present the formal description of a {\em queue} as Object Type~\ref{type:queue}.

\begin{figure}[h]\fbox{\begin{minipage}\textwidth		
\begin{type}[Queue]
A queue of elements from $\N^+$ is described as follows:
\label{type:queue}
\begin{itemize}[parsep=0pt,partopsep=0pt]
    \item 
    $\Sigma = \bigcup_{n \in \N} (\N^+)^n$ 
    \item
    $OP = \{\Enqueue, \Dequeue\}$
    \item
    $ARG_{\Enqueue} = \N^+, ARG_{\Dequeue} = \{\bot\}$.
    \item
    $RES = \{\ack \} \cup \N^+$ 
    \item
    Transition function $\delta$ is defined as follows:
    \begin{itemize}
        \item 
        $\delta(\sigma, \pi, \Enqueue, v) = (\sigma \circ v, \ack)$
        \item
        $\delta(\sigma, \pi, \Dequeue, \bot) = (tail(\sigma), head(\sigma))$ where $\sigma \neq ()$, the empty sequence.
    \end{itemize}
\end{itemize}
\end{type}
\end{minipage}}\end{figure}

\begin{definition}[algorithm]
    A {\em (concurrent) algorithm} is a tuple $(\Pi, \Omega, \mathcal{C}_0)$, where:
    \begin{itemize}
        \item 
        $\Pi$ is a set of processes, where each process $\pi \in \Pi$ has a program and private registers, including a program counter $pc_\pi$ which points to the line to be executed in the program.
        A {\em process's state} at any point in time is described by the values of its private registers.

        \item
        $\Omega$ is a set of objects; each object has a type, and is in one of its states at any point in time.

        \item
        $\mathcal{C}_0$ is a non-empty set of configurations, called {\em initial configurations}, where a {\em configuration} is an assignment of a state to each object $\omega \in \Omega$ and an assignment of values to the private registers of each process $\pi \in \Pi$.
    \end{itemize}
\end{definition}

\begin{definition}[step, event, run]
\mbox{ }
    \begin{itemize}
    \item
    A {\em step} of an algorithm is a triple $(C, (\pi, \ell), C')$ such that $C$ is a configuration, $\pi$ is a process, $\ell$ is the line of code pointed to by $\pi$'s program counter in $C$, and $C'$ is a configuration that results when $\pi$ executes line $\ell$ from $C$.
    
    \item
    The {\em event} corresponding to a step $(C, (\pi, \ell), C')$ is $(\pi, \ell)$, i.e., process $\pi$ executing line $\ell$.
    
    \item
    A {\em run} of an algorithm is a finite sequence $C_0, (\pi_1, \ell_1), C_1,(\pi_2,\ell_2),C_2,\ldots,(\pi_k,\ell_k),C_k$ or an infinite sequence $C_0, (\pi_1, \ell_1), C_1,(\pi_2,\ell_2),C_2,\ldots$ such that $C_0$ is an initial configuration and each triple $(C_{i-1},(\pi_i, \ell_i),C_i)$ is a step.
    
    \end{itemize}
\end{definition}

\subsection{Implementation of an Object}

Implementing complex objects, such as queues and snapshots, from primitive objects supported by the underlying hardware (registers supporting read, write, CAS etc.) is a central problem in multiprocessor programming.
Below, we describe what an implementation entails. 
Later on, we will define what it means for an implementation to be {\em correct}, in the sense of linearizability.

\begin{definition}[implementation]
	An {\em implementation $\Ocal$ of an object of type $\tau$ initialized to state $\sigma_0$} for a set of processes $\Pi$ specifies
	\begin{itemize}
		\item 
		A set of objects $\Omega$ called the {\em base objects} along with their types and initial states.
		
		\item
		A set of procedures $\Ocal.op_{\pi}(arg)$ for each $\pi \in \Pi$, $op \in \tau.OP$, and $arg \in \tau.ARG_{op}$.
		The objects accessed in the code of the procedures must all be in $\Omega$.
	\end{itemize}

	To execute an operation $op$ with argument $arg$ on the implemented object $\Ocal$, a process $\pi$ invokes the method $O.op_{\pi}(arg)$ (and executes the code in the procedure).
	The value returned by the method is deemed $\Ocal$'s response to this operation invocation.
\end{definition}

Notice that this definition only captures the syntactic aspect of an implementation.
We will now build up to defining the correctness of an implementation.

Intuitively, the implemented object is correct if it behaves like an atomic object of the same type.
To formally capture correctness, we define {\em behavior}, the notion of an {\em atomic implementation}, and the correctness condition of {\em linearizability}.

\subsection{Behaviors of an Implementation}

Consider an object implementation $\Ocal$, and a run $R$ in which processes invoke operations on $\Ocal$, execute the corresponding procedures of $\Ocal$, and receive responses.
By the definition of a run, $R$ is an alternating sequence of configurations and events.
Some of the events are {\em invocation events}, i.e. calls to $\Ocal$'s procedures, and some are {\em response events}, i.e. the execution of return statements of $\Ocal$'s procedures.
(Of course, there are other events, such as the execution of other lines between the call and return of a procedure.)
We call the subsequence of $R$ that includes only the invocation and response events the {\em behavior} in $R$.
For example, if $\Ocal$ is an initially empty queue, a behavior can be 
\begin{equation*}
\scriptstyle
(\pi_1, \text{\bf invoke } \Enqueue_{\pi_1}(5)), (\pi_2, \text{\bf invoke } \Dequeue_{\pi_2}()), (\pi_3, \text{\bf invoke } \Enqueue_{\pi_3}(7)), (\pi_2, \text{\bf response } 7), (\pi_2, \text{\bf invoke } \Enqueue_{\pi_2}(9))
\end{equation*}
Every possible behavior of an implementation $\Ocal$ can be generated by the algorithm of Figure~\ref{alg:generator}, where each process repeatedly chooses an operation non-deterministically, invokes it by calling the corresponding procedure, and executes the procedure until it returns (receives a response).
The next definition captures this discussion.

\begin{figure}[h]
	\fbox{\begin{minipage}\textwidth
			\begin{algorithmic}[1]
				\medskip
				\Statex {\bf Initial Configurations:}
				\begin{itemize}
					\item 
					$\omega$ is an object of type $\tau$, in its initial state $\sigma_0$.
					\item
					Each process $\pi \in \Pi$ is assigned the program $main_\pi()$;
					i.e. $pc_\pi$ is initialized to line $a$.
					\item
					Every other private register of each $\pi \in \Pi$ is initialized arbitrarily.
				\end{itemize}
				
			\end{algorithmic}

   			\medskip 

			\begin{algorithmic}[1]				
				\Statex \cmdprogram $main_\pi()$
				
				\let\fmtlinenumber\mathalph 
				
				\State \tab \cmdwhile $\true$ \cmddo
				choose $(op, arg) \in \{(o, a) \mid o \in \tau.OP, a \in \tau.ARG_o \}$ and invoke $\Ocal.op_\pi(arg)$
				
				\let\fmtlinenumber\arabic 

			\end{algorithmic}
   				\caption{ 
                    {\em Generator algorithm} $\A(\Ocal)$ that generates all behaviors of an implemented object $\Ocal$ of type $\tau$.
                    Code shown for process $\pi \in \Pi$.
				}	
    	    \label{alg:generator}			

	\end{minipage}}
\end{figure}

\begin{definition}[implementation runs and behaviors]
	Let $\Ocal$ be an implementation of a type $\tau$ initialized to $\sigma_0$ for a set $\Pi$ of processes.
	We define the {\em runs} of $\Ocal$ to be the set of all runs of the {\em generator algorithm}, $\A(\Ocal)$, presented in Figure~\ref{alg:generator}.
	Let $\Rcal$ be the set of all runs of $\Ocal$.
	For any run $R \in \Rcal$, we define $behavior(R)$ to be the subsequence of all the invocation and response events in $R$.
	The set of all {\em behaviors} of $\Ocal$ is $\{behavior(R) \mid R \in \mathcal{R}\}$.
\end{definition}

\subsection{The Atomic Implementation}

Implementing an object from base objects of other types is often challenging, but implementing an object $\Ocal$ from a base object $\omega$ of the same type is trivial: each procedure $\Ocal.op_\pi(arg)$ is implemented simply by executing $\omega.op_\pi(arg)$ and returning the received response.
We call this implementation the {\em atomic implementation}.

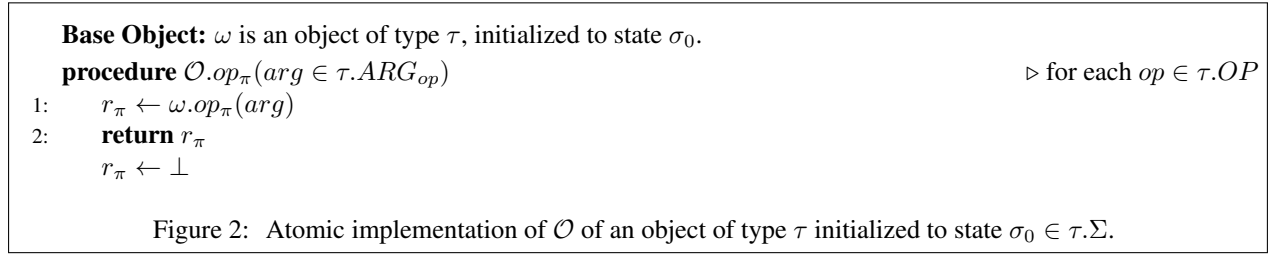
\begin{figure}[h]\fbox{\begin{minipage}\textwidth
			
			\begin{algorithmic}[1]
				\Statex
				\Statex {\bf Base Object:}  $\omega$ is an object of type $\tau$, initialized to state $\sigma_0$.
			\end{algorithmic}
			
			\begin{algorithmic}[1]	
				\Statex \cmdprocedure $\Ocal.op_\pi(arg \in \tau.ARG_{op})$ \Comment{for each $op \in \tau.OP$}
				\State \tab $r_\pi \gets \omega.op_\pi(arg)$
				\State \tab \cmdreturn $r_\pi$
			    \Statex\tab $r_\pi \gets \bot$
			\end{algorithmic}
			
			\caption{ 
				Atomic implementation of $\Ocal$ of an object of type $\tau$ initialized to state $\sigma_0 \in \tau.\Sigma$.
			}
\label{implementation:atomic}
\end{minipage}}
\end{figure}

\begin{definition}[atomic implementation]
	The {\em atomic implementation} of an object $\Ocal$ of type $\tau$, initialized to $\sigma_0$, is the implementation presented in Figure~\ref{implementation:atomic}. 
	(On line 2, the implementation resets $r_\pi$ to $\bot$ as soon as it returns the value.)
\end{definition}

\subsection{Linearizability}

We are now ready to define {\em linearizability}.
Intuitively, an object implementation is linearizable if it behaves like an atomic object of the same type.
Formally:

\begin{definition}[linearizability]
	For a set $\Pi$ of processes,
	let $\Ocal$ be an implementation of an object of type $\tau$ initialized to $\sigma_0$, and let $\Ocal_{atomic}$ be the atomic implementation of an object of type $\tau$ initialized to $\sigma_0$.
	Furthermore, let $\Rcal$ be the set of all runs of $\A(\Ocal)$ and $\Rcal_{atomic}$ be the set of all runs of $\A(\Ocal_{atomic})$.
	We say a run $R_{atomic} \in \Rcal_{atomic}$ is a {\em linearization} of a run $R \in \Rcal$ if $behavior(R) = behavior(R_{atomic})$.
	Correspondingly, we say that a run $R \in \Rcal$ is {\em linearizable} if it has a linearization $R_{atomic} \in \Rcal$; equivalently, if $behavior(R)$ is also a behavior of $\Ocal_{atomic}$.
	We say that the implementation $\Ocal$ is linearizable if every finite run $R \in \Rcal$ is linearizable.
	Equivalently, $\Ocal$ is linearizable if every finite behavior of $\Ocal$ is a behavior of $\Ocal_{atomic}$.
\end{definition}

\subsection{Strong Linearizability}

In general, a run $R$ of a linearizable implementation can have multiple linearizations.
Intuitively, a linearizable object implementation satisfies {\em strong linearizability} if for any run of the implementation $R$ the object can ``commit to a specific linearization'' $\Lcal(R)$, such that the linearization of any extension of the run $R$ is an extension of $\Lcal(R)$.

\begin{definition}
For a set $\Pi$ of processes,
let $\Ocal$ be an implementation of an object of type $\tau$ initialized to $\sigma_0$, and let $\Ocal_{atomic}$ be the atomic implementation of an object of type $\tau$ initialized to $\sigma_0$.
Furthermore, let $\Rcal$ be the set of all runs of $\A(\Ocal)$ and $\Rcal_{atomic}$ be the set of all runs of $\A(\Ocal_{atomic})$.
An implementation is {\em strongly linearizable}, if there is a {\em linearization function} $\Lcal: \Rcal \to \Rcal_{atomic}$ that maps each run $R$ of the implementation to an atomic run $\Lcal(R)$ of the atomic implementation that is a linearization of it, such that if $R_{pre}$ is a prefix of $R$, then $\Lcal(R_{pre})$ is a prefix of $\Lcal(R)$.
\end{definition}

Not all linearizable implementations are strongly linearizable, since, for some implementations, we need to extend different linearizations of a run $R$ to linearize different extensions of $R$.
This notion of {\em strong} linearizability is subtle, but has been shown important in preserving hyperproperties of composed algorithms, such as output probability distributions \cite{StrongLinearizability, Attiya19}.

\section{Our Proof Technique for Linearizability}

Our goal is to devise a scheme by which algorithmists can produce machine verifiable proofs that complex concurrent object implementations are linearizable.
To prove linearizability of an implementation $\Ocal$, our technique is to augment $\Ocal$'s code to produce an {\em augmented implementation} $\Oaug$ in such a way, that $\Ocal$ is linearizable if and only if the generator of $\Oaug$, i.e. $\A(\Oaug)$, satisfies a certain simple invariant.
In the following, we describe the rules for augmentation, and some specific augmentations of interest.

\subsection{Augmenting an Implementation}

Informally, an {\em augmented implementation} $\Oaug$ has all the code of $\Ocal$, and more.
In particular, $\Oaug$ may employ additional {\em auxiliary} base objects, and at each line manipulate these auxiliary objects along with performing the code of that line in the original implementation. 

\begin{definition}[augmentation]
	Let $\Ocal$ be an implementation of type $\tau$ with initial state $\sigma_0$ from base objects $\Omega$ for processes $\Pi$.
	An {\em augmentation} $\Oaug$ of $\Ocal$ is also an implementation of type $\tau$ with initial state $\sigma_0$ for processes $\Pi$, with the following properties:
	\begin{itemize}
		\item 
		$\Oaug$ may employ an additional set of {\em auxiliary} base objects $\Omega_{aux}$; thus, $\Oaug$ is an implementation from $\Omegaaug = \Omega \cup \Omega_{aux}$.
		
		\item 
		For each line $\ell$ of each procedure $\Ocal.op_\pi(arg)$,
		the augmented procedure has a bijectively corresponding line $\ellaug$ of $\Oaug.op_\pi(arg)$, such that; $\ellaug$ contains all the code of the corresponding line $\ell$ and, optionally, some additional code that only changes objects in $\Omega_{aux}$.
	\end{itemize}
\end{definition}

Since the additional code has no impact on the original base objects of $\Ocal$ or the private registers of any $\pi \in \Pi$ (including $pc_\pi$), we note that $\Oaug$ and $\Ocal$ have identical behaviors as summarized below.

\begin{observation}
\label{obs:augmentation}
	If $\Oaug$ is an augmentation of $\Ocal$, then the set of behaviors of $\Oaug$ is identical to the set of behaviors of $\Ocal$.	
\end{observation}

When we have two implementations $\Ocal$ and $\Oaug$, we will often be interested in analyzing {\em coupled runs} of these implementations as defined below:

\begin{definition}[coupled runs]
    Let $\Ocal$ and $\Oaug$ be implementations of the same object type $\tau$ for the same set of processes $\Pi$, with bijectively corresponding lines.
    We say a run $R_k = C_0, (\pi_1, \ell_1), C_1, (\pi_2,\ell_2),\ldots,C_k$ of $\A(\Ocal)$ and a run $\Raug_k = \Caug_0,(\pi_1, \ellaug_1), \Caug_1, (\pi_2, \ellaug_2),\ldots,\Caug_k$ of $\A(\Oaug)$ are coupled,
    if $C_0$ and $\Caug_0$ share an initialization for all their common variables, and each $\ell_i$ and $\ellaug_i$ are corresponding lines for each $1 \le i \le k$. 
\end{definition}

\subsection{The Full Tracker Augmentation}

In this section, we consider an implementation $\Ocal$ of an object type $\tau$, and we define a specific augmentation $\Ofull$ called the {\em full tracker} of $\Ocal$, which aims to keep track of all possible linearizations of a run $R$ of $\Ocal$, as the run unfolds.
Recall that by definition, a linearization $R_{atomic}$ of a run $R$ is a run of the atomic implementation $\Ocal_{atomic}$ of type $\tau$ whose behavior matches that of $R$.
In practice, $\Oaug$ will actually maintain just the final configuration $C_{atomic}$ of each linearization $R_{atomic}$, rather than the whole linearization, thus we now take a closer look at these atomic configurations.

After any run $R_{atomic}$, the final configuration can be characterized by the state $\sigma \in \tau.\Sigma$ of the atomic object, and the states of each of the processes $\pi \in \Pi$.
For each process $\pi$, there are three types of possible states:
\begin{enumerate}
    \item 
    either $\pi$ is {\em idle}, meaning $\pi$ has no currently invoked but un-returned operation, i.e. $pc_\pi$ points at the single line (line~$a$) of the generator algorithm and $r_\pi = \bot$.
    \item
    or $\pi$ has invoked an operation $oper_\pi(argu)$ which is {\em yet-to-linearize}, i.e. $pc_\pi$ points to line~1 of procedure $oper$'s code and $r_\pi = \bot$.
    \item
    or $\pi$ {\em has linearized} its operation $oper_\pi(argu)$ with some response $resp$, i.e. $pc_\pi$ points to line~2 of procedure $oper$'s code and $r_\pi = resp$.
\end{enumerate}
We capture the states of the processes with a function $f(\pi) = (f(\pi).op, f(\pi).arg, f(\pi).res)$, which maps each process $\pi$ to a triple, where
\begin{enumerate}
    \item 
    if $\pi$ is idle, then $f(\pi) = (\bot,\bot,\bot)$
    \item
    if $\pi$ has invoked an operation $oper_\pi(argu)$ which is yet-to-linearize, then $f(\pi)= (oper, argu, \bot)$
    \item
    and if $\pi$ has linearized its operation $oper_\pi(argu)$ with some response $resp$, then $f(\pi) = (oper, argu, resp)$
\end{enumerate}

Thus, we identify each atomic configuration $C_{atomic}$, i.e. configuration of $\A(\Ocal_{atomic})$, as an object-state, process-states pair $(C_{atomic}.\sigma, C_{atomic}.f)$ as described above.
Here on, we will refer to these pairs as the atomic configurations.

For the full tracker, we augment the implementation $\Ocal$ with an object $\Mfull$ that stores the set of all final configurations of linearizations of the implemented object's run thus far.
We call $\Mfull$ the {\em meta-configuration} of the tracker, since it ``contains all the atomic configurations in which a corresponding atomic run can be in''. 
The meta-configuration set is, in fact, the only auxiliary object we use in the tracker augmentation, thus $\Omega_{aux} = \{\Mfull\}$.

Next, we complete the description of $\Ofull$ by describing how the meta-configuration $\Mfull$ is initialized and manipulated by the full tracker.
Initially, all processes are idle, and $\Ocal$ is in its initial state $\sigma_0$; thus we initialize $\Mfull = \{(\sigma_0, f_0)\}$, where $f_0(\pi) = (\bot,\bot,\bot)$ for all process $\pi \in \Pi$.

There are three types of lines in a procedure $\Ocal.op_{\pi}(arg)$: {\em invocation}, {\em return}, and {\em intermediate lines}.
The augmenting code at each line $\ell$ updates $\Mfull$, based on $\ell$'s type. 
Specifically,
\begin{itemize}
	\item
	\underline{Case: $\ell$ is the invocation of $op_{\pi}(arg)$}.
	We will maintain the invariant that $\Mfull$ stores all final configurations of linearizations of the implemented object's current run.
	Thus, every configuration $C \in \Mfull$ will reflect that $\pi$ is idle before the invocation.
	In the augmentation to the invocation line, we will update each $C \in \Mfull$ to a $C'$ which reflects that $\pi$ invokes a pending operation $op$ with argument $arg$, and then further evolve $C'$ to reflect that any arbitrary subset of processes with pending operations (possibly including $\pi$) can linearize after $\pi$ invokes its operation.
	Before formally stating the augmenting code to the invocation line, we develop some helpful notation.
	
	\begin{itemize}
		\item 
		For an atomic configuration $C = (C.\sigma, C.f)$ in which $\pi$ is idle, we define:
		\begin{equation*}
			invoke(C, \pi, op_{\pi}(arg)) \triangleq C'
		\end{equation*}
		where $C'.\sigma = C.\sigma$, $C'.f(\pi) = (op, arg, \bot)$, and for each $\bar\pi \ne \pi$, $C'.f(\bar\pi) = C.f(\bar\pi)$.
		
		\item 
		For an atomic configuration $C$, and a process $\pi \in \Pi$, we define the predicate
		\begin{equation*}
			pending(\pi, C) \equiv C.f(\pi).op \ne \bot \wedge C.f(\pi).res = \bot
		\end{equation*}
		to capture whether $\pi$ has a pending operation in $C$. 
		We further define
		\begin{equation*}
			pending(C) \equiv \{\pi \in \Pi \mid pending(\pi, C)\}
		\end{equation*}
		to be the set of all processes with pending operations in $C$.
		
		\item 
		For any $S \subseteq pending(C)$ of processes with pending operations, and a permutation of those processes $\alpha = (\alpha_1,\ldots,\alpha_{|S|}) \in Perm(S)$, we define the transition function $\delta^*(C, \alpha)$ recursively below.
		(Informally, $\delta^*(C, (\alpha_1,\ldots,\alpha_k)) = C'$ if applying the pending operations of processes $\alpha_1,\ldots,\alpha_k$, in that order, updates the configuration $C$ to $C'$.)
		\begin{itemize}
			\item[$\bullet$]
			$\delta^*(C, ()) = C$, where $()$ is the empty sequence
			\item[$\bullet$]
			$\delta^*(C, (\alpha_1,\ldots,\alpha_k)) = \delta^*(C', (\alpha_2,\ldots,\alpha_k))$, where 
			\begin{align*}
			\exists r \in RES: \hspace{0.2in}	&(C'.\sigma, r) = \delta(C.\sigma, \alpha_1, C.f(\alpha_1).op, C.f(\alpha_1).arg) \\
					  		&\wedge\: \forall \pi \ne \alpha_1, C'.f(\pi) = C.f(\pi) \\
						  	&\wedge\: C'.f(\alpha_1)= (C.f(\alpha_1).op, C.f(\alpha_1).arg, r)
			\end{align*}

		\end{itemize}  
	\end{itemize}
	
	Formally, the augmenting code at line $\ell$ is:
	$\Mfull \gets EvolveInv(\Mfull, op_\pi(arg))$, where $EvolveInv$ is defined by
	\begin{equation}
  	EvolveInv(\M, op_\pi(arg)) \triangleq \left\{ C''\ \middle\vert	
  	            \begin{array}{l}
    	                \exists C \in \M, C': \\
    					    \tab C' = invoke(C, \pi, op_\pi(arg)) \: \wedge \\
							\tab  \exists S \subseteq pending(C'), \alpha \in Perm(S): C'' = \delta^*(C', \alpha)
  				\end{array}
  			 \right\}
	\end{equation} 
	We say {\em configuration $C''$ results from $\pi$ invoking $op_\pi(arg)$ and $\alpha_1,\ldots,\alpha_k$ linearizing after configuration $C$}.

	\item
	\underline{Case: $\ell$ is an intermediate line of $op_{\pi}(arg)$}.
	The augmenting code for this line reflects that any time any process executes an intermediate line, it presents an opportunity for an arbitrary subset of pending operations to linearize in an arbitrary order.
	Formally, the augmenting code at line $\ell$ is:
	$\Mfull \gets Evolve(\Mfull)$, where $Evolve$ is defined by
	\begin{equation}
	Evolve(\M) \triangleq \{C' \mid \exists C \in M, S \subseteq pending(C), \alpha \in Perm(S): C' = \delta^*(C, \alpha)\}
	\end{equation} 
	We say {\em configuration $C'$ results from $\alpha_1,\ldots,\alpha_k$ linearizing after configuration $C$}.

	\item
	\underline{Case: $\ell$ is a $\cmdreturn \: res$ statement from a procedure $op_{\pi}(arg)$}.
	Those atomic configurations that do not reflect that the operation has linearized with a response of $res$ are no longer tenable, and thus these are filtered out of $\Mfull$.
	On the other hand, the configurations that show this response of $res$ are retained and updated to reflect that $\pi$ becomes idle after the return.
	Furthermore, any arbitrary subset of pending processes can linearize after $\pi$ returns.
	Before formally stating the augmenting code to the invocation line, we develop some helpful notation.
	\begin{itemize}
	\item 
	For an atomic configuration $C = (C.\sigma, C.f)$ in which $\pi$ has linearized its operation $op_\pi(arg)$ with return value $res$, we define:
	\begin{equation*}
		return(C, \pi, res) \triangleq C'
	\end{equation*}
	where $C'.\sigma = C.\sigma$, $C'.f(\pi) = (\bot, \bot, \bot)$, and for each $\bar\pi \ne \pi$, $C'.f(\bar\pi) = C.f(\bar\pi)$.
    \end{itemize}
	Formally, the augmenting code at line $\ell$ is: $\Mfull \gets EvolveRet(\Mfull, \pi, res)$, where $EvolveRet$ is defined by

	\begin{equation}
  	EvolveRet(\M, \pi, res) \triangleq \left\{ C''\ \middle\vert	
        \begin{array}{l}
                \exists C \in \M, C': \\
                    \tab C.f(\pi).res = res \: \wedge \\
                    \tab C' = return(C, \pi, res) \: \wedge \\
                    \tab  \exists S \subseteq pending(C'), \alpha \in Perm(S): C'' = \delta^*(C', \alpha)
        \end{array}
     \right\}
	\end{equation} 
	We say {\em configuration $C''$ results from $\pi$ returning $res$ and $\alpha_1,\ldots,\alpha_k$ linearizing after configuration $C$}.
\end{itemize}

\begin{definition}[full tracker]
	For any implementation $\Ocal$, $\Ofull$ is the {\em full tracker} as specified above with the single auxiliary {\em meta-configuration} variable $\Mfull$.
\end{definition} 

\subsection{Main Theorem}

The crafting of the tracking augmentation yields a powerful technique for producing {\em machine verified proofs} of linearizability.
This technique falls out from the main theorem of this section, which reduces the complex question of whether an implementation $\Ocal$ is linearizable to the verification of a simple invariant.
In particular, $\Ocal$ is linearizable {\em if and only if} $\Mfull \ne \nullset$ is an invariant of $\A(\Ofull)$.

Exploiting the {\em if} direction of this theorem, we can obtain machine verified proofs of linearizability by proving the invariant in a software system such as TLAPS (Temporal Logic of Actions Proof System) or Coq.
The {\em only if} direction of the theorem assures us that this proof technique is complete.

\begin{restatable}{theorem}{maintheorem} \label{maintheorem}
	Let $\Ocal$ be an implementation of an object of type $\tau$ initialized to state $\sigma_0$ for a set of processes $\Pi$,
	$\Ocal$ is linearizable {\em if and only if} $\Mfull \ne \nullset$ is an invariant of $\A(\Ofull)$.
 \end{restatable}
Recall that $\Ofull$ is the full tracker of $\Ocal$ with meta-configuration $\Mfull$, and $\A(\Ofull)$ is the algorithm from Figure~\ref{alg:generator} in which processes repeatedly make calls to procedures of $\Ofull$.
The remainder of the section develops the necessary machinery and proves Theorem~\ref{maintheorem}.


Let $\Ocal$ be an implementation of an object of type $\tau$ initialized to $\sigma_0$ for a set of processes $\Pi$, and
let $O$ be an atomic object with the same type and initialization, and for the same set of processes.
Finally, recall that $\Ofull$ is the full tracker of $\Ocal$ with meta-configuration variable $\Mfull$. 

For a generic variable $V$ of any given implementation $I$, and a run $R_I$ of the generator $\A(I)$, we let $V(R_I)$ denote the value of $V$ in the final configuration of $\A(I)$.
We will be particularly interested in the meta-configurations $\Mfull(\Rfull)$ for runs $\Rfull$ of $\A(\Ofull)$.

By Observation~\ref{obs:augmentation}, $\Ocal$ is linearizable if and only if the finite behaviors of $\Ofull$ are behaviors of $O$.
Our strategy to prove the main theorem thereby, will be to show that for any run $\Rfull$ of $\Ofull$, the set of atomic runs sharing its behavior will have exactly the final configurations $\Mfull(\Rfull)$.

To express the above statement, we develop a bit more notation.
For any behavior $B$, we let $AtomicRuns(B)$ be the set of atomic runs that exhibit behavior $B$.
For any run $\Rfull$ of the implementation, we define $Linearizations(\Rfull) \triangleq AtomicRuns(behavior(\Rfull))$ to be the set of all possible linearizations of $\Rfull$.
Finally, for run $R$, we let $\C(R)$ denote the final configuration of $R$, and for a set of runs $\Rcal$, we let $\C(\Rcal) \triangleq \{\C(R) \mid R \in \Rcal \}$ denote the set of final configurations of all those runs.

We are now ready to prove the key lemma, which states that the meta-configuration variable tracks, precisely, the set of final configurations of linearizations of an implementation run. 

\begin{restatable}{lemma}{mainlemma} \label{lem:mainlemma}
	Let $\Rfull$ be any finite run of $\Ofull$, then the following equality holds:
	\begin{equation*}
		\Mfull(\Rfull) = \C(Linearizations(\Rfull))
	\end{equation*}	
\end{restatable}


\begin{proof}
    We present the full formal inductive proof of this lemma in Appendix~\ref{sec:proof-of-full-tracker-main-lemma}.
\end{proof}


\maintheorem*

\begin{proof}
	We will prove the theorem by showing the forward and reverse directions separately:
	\begin{itemize}
		\item 
		For the forward direction:
		assume $\Mfull \ne \nullset$ is an invariant of $\A(\Ofull)$.
		Thus, for any finite run $\Rfull$ of $\A(\Ofull)$, we note $\Mfull(\Rfull)$ is non-empty.
		So, by Lemma~\ref{lem:mainlemma}, there is a linearization $L$ of $\Rfull$.
		Since, an arbitrary finite run $\Rfull$  of $\A(\Ofull)$ is linearizable, we conclude that $\Ofull$ is linearizable.
		Since $\Ofull$ has the same behaviors as $\Ocal$, thus $\Ocal$ is linearizable.
		
		\item
		For the reverse direction:
		assume that $\Ocal$ is linearizable.
		This implies $\Ofull$ is linearizable.
		So, every finite run $\Rfull$ of $\A(\Ofull)$ has at least one linearization $L$.
		Thus, by Lemma~\ref{lem:mainlemma}, $\Mfull(\Rfull)$ is non-empty.
		Since, $\Mfull(\Rfull)$ is non-empty for an arbitrary finite run $\Rfull$ of $\A(\Ofull)$, we conclude that $\Mfull \ne \nullset$ is an invariant of $\A(\Ofull)$.
	\end{itemize}
\end{proof}

\section{(Partial) Trackers}

The proof of Theorem~\ref{maintheorem} demonstrates that, for an implementation $\Ocal$, the full tracker $\Ofull$, in essence, tracks ``all possible linearizations'' in its meta-configuration.
To show that $\Ocal$ is linearizable however, it suffices to show the existence of a single linearization.
That is, we need not track {\em all} linearizations, but just ensure that we track {\em at least one}.
In practice, tracking a subset of linearizations can be easier for a prover, especially one who knows the structure of the implementation well.
Thus, in general, we will be interested in a (partial) tracker, whose meta-configuration $\Mpart$ maintains a subset of the full meta-configuration $\Mfull$.
We define such trackers.

\begin{definition}[trackers]
	Let $\Ocal$ be an implementation of a type $\tau$ in the initial state $\sigma_0$ for a set of processes $\Pi$.
	An augmentation $\Opart$ is a {\em (partial) tracker} of $\Ocal$, if the following conditions are met:
	\begin{itemize}
		\item
		The auxiliary variable set $\Omega_{aux}$ of $\Opart$ contains a meta-configuration variable $\Mpart$ that holds a set of atomic configurations.
		($\Omega_{aux}$ may also contain other additional variables.)
		
		\item
		$\Mpart$ is initialized to $\{(\sigma_0, f_0)\}$, where $f_0(\pi) = (\bot, \bot, \bot)$ for every $\pi \in \Pi$.
		
		\item
		For each type of line $\ell$ (whether invocation, intermediate, or return) in the original algorithm, the augmented line $\ellpart$ updates the tracker to ensure the new value of $\Mpart$ is a subset of the configurations that would arise by evolving $\Mpart$ according to the rules of the full tracker.
		That is, if the update rule of the tracker at line $\ellpart$ is $\Mpart \gets Evolve_\ell(\Mpart)$, then:
		\begin{itemize}
		 	\item
        	\underline{Case: $\ell$ is the invocation of $op_{\pi}(arg)$}.
        	\begin{equation}
          	Evolve_\ell(\Mpart) \subseteq EvolveInv(\Mpart, op_\pi(arg))
        	\end{equation} 
        
        	\item
        	\underline{Case: $\ell$ is an intermediate line of $op_{\pi}(arg)$}.
        	\begin{equation}
        	Evolve_\ell(\Mpart) \subseteq Evolve(\Mpart)
        	\end{equation} 

        	\item
        	\underline{Case: $\ell$ is a $\cmdreturn \: res$ statement from a procedure $op_{\pi}(arg)$}.
        	\begin{equation}
        	Evolve_\ell(\Mpart) \subseteq EvolveRet(\Mpart, \pi, res)
        	\end{equation} 
		 \end{itemize}	
        
	\end{itemize} 
\end{definition} 

\begin{theorem}
\label{thm:partial-tracker}
	Let $\Ocal$ be an implementation of an object of type $\tau$ initialized to state $\sigma_0$ for a set of processes $\Pi$,
	and $\Opart$ be any partial tracking augmentation of $\Ocal$ with meta-configuration variable $\Mpart$.
	If $\Mpart \ne \nullset$ is an invariant of $\A(\Opart)$, then $\Ocal$ is linearizable.
\end{theorem}

\begin{proof}
	Let $\Mfull$ be the meta-configuration of the full tracker $\Ofull$.
	Consider any run $\Rfull = \Cfull_0, (\pi_1, \ellfull_1), \Cfull_1, (\pi_2, \ellfull_2), \ldots$ of $\A(\Ofull)$, and recall that we say the run $\Rpart = \Cpart_0, (\pi_1, \ellpart_1), \Cpart_1, (\pi_2, \ellpart_2), \ldots$ of $\A(\Opart)$ is coupled with $\Rfull$, if $\Cpart_0$ has the same initialization for the private registers and objects in $\tau.\Omega$, and for each $i$, $\ellpart_i$ is the line in $\Opart$ that corresponds to $\ellfull_i$ in $\Ofull$, i.e., these are augmentations of corresponding lines in $\Ocal$.
	Since $\Ofull$ and $\Opart$ augment the same implementation, all runs of $\A(\Ofull)$ and $\A(\Opart)$ are coupled as $(\Rfull, \Rpart)$ in this way.  
	For coupled runs $\Rfull$ and $\Rpart$, by definition of a partial tracking augmentation, it is clear that $\Mpart \subseteq \Mfull$ throughout the coupled run.
	Thus, $\Mpart \ne \nullset$ being an invariant of $\Opart$ implies $\Mfull \ne \nullset$ is an invariant of $\Ofull$.
	So, by Theorem~\ref{maintheorem}, the proof is complete.
\end{proof}

\section{Proving Strong Linearizability}

In the previous section, we motivated partial trackers as simpler ways to prove standard linearizability.
In this section, we will show how to use partial trackers to obtain a {\em sound and complete} method for proving {\em strong linearizability}.

The pivotal difference between strong linearizability and (standard) linearizability, is that the former consistency condition requires that we should be able to commit to a single linearization $\Lcal(R)$ for any run $R$, such that the linearization of every extension of $R$ will be an extension of $\Lcal(R)$.
Our key insight lies here.
The full tracker's meta-configuration $\Mfull$ maintains the final configurations of {\em every possible linearization} of $R$ hoping to extend whichever ones may work as the future of the run unfolds; thus, it effectively ensures that ``no linearization will be missed'' thereby yielding a complete method for proving linearizability.
However, informally speaking, by maintaining every possible linearization, it does exactly the opposite of {\em committing to single linearization}, which is the key to showing strong linearizability.

Our idea therefore, is to demonstrate strong linearizability by demonstrating a partial tracker $\Opart$ whose meta-configuration $\Mpart$ holds precisely one configuration at any point in time.
The intuition is that maintaining a single configuration $\Mpart(\Rpart_k) = \{C_k\}$ at the end of a run $\Rpart_k$ of length $k$ is akin to committing to a single linearization $L_k$ (with $\C(L_k) = C_k$) for this run, and that this unique linearization is getting extended as the run gets extended.
Of course, there is a small catch.
The configuration $C_k$ could be the final configuration of many different linearizations, which all happen to have the same final configuration.
Nevertheless, we will resolve this hiccup by proving that a cleverly chosen particular one of these linearizations can be picked as $\Lcal(\Rpart_k)$.

\begin{restatable}{theorem}{thmstronglinearizability} \label{thmstronglinearizability}
	Let $\Ocal$ be an implementation of an object of type $\tau$ initialized to state $\sigma_0$ for a set of processes $\Pi$,
	$\Ocal$ is strongly linearizable {\em if and only if} there exists a partial tracker $\Opart$ such that $\A(\Opart)$ satisfies the invariant $|\Mpart| = 1$.
	
\end{restatable}

\begin{proof}
    We present the full inductive proof of this theorem in Appendix~\ref{sec:strong-linearizability-theorem}.    
\end{proof}

\section{Applying the Technique}
\label{sec:example}

We have used our tracking technique to produce machine-certified proofs of three concurrent data structures: we proved the linearizability of the Herlihy-Wing queue and Jayanti's single-writer single-scanner snapshot; and we have proved the strong linearizability of the Jayanti-Tarjan union-find object.
The Herlihy-Wing queue implementation, presented in Figure~\ref{impl:queue}, is particularly notorious for being difficult to prove correct, so we will use that data structure as a running example in this section.
We present the other two results in Appendices~\ref{sec:snapshot} and \ref{sec:union-find}, respectively.

This section has two goals:
(1) describing the {\em proof process}, i.e., the steps taken by a {\em human-prover} in order to produce a machine-certified proof of linearizability; and
(2) describing the {\em verification process}, i.e., the steps required of the {\em human-verifier} in order to confirm the proof's correctness.
Overall, with the tracking technique, the prover does work comparable to what would be done in a pen-and-paper proof.
The verifier's job on the other hand, becomes very simple and fast.

\begin{figure}[H]\fbox{\begin{minipage}\textwidth		
\begin{algorithmic}[1]
    \medskip
    \Statex {\bf Base Objects:}
    \begin{itemize}[noitemsep,topsep=0pt,parsep=0pt,partopsep=0pt]
        \item $X$ is a F\&Inc/read register initialized to 1.
        \item $Q[1, 2, \ldots]$ is an infinite FAS/read/write array, where each $Q[i]$ is initialized to $\bot$.
    \end{itemize}
\end{algorithmic}

\begin{multicols}{2}
\begin{algorithmic}[1]
    \medskip
    \State \cmdprocedure $\Ocal.\Enqueue_{\pi}(v_{\pi})$
    \State \tab $i_{\pi} \gets \FAInc{X}$
    \State \tab $Q[i_{\pi}] \gets v_{\pi}$
    \State \tab \cmdreturn $\ack$
\end{algorithmic}

\columnbreak

\begin{algorithmic}[1]
\setcounter{ALG@line}{4}
    \State \cmdprocedure $\Ocal.\Dequeue_{\pi}()$
    \State \tab $l_{\pi} \gets X$
    \State \tab \cmdif $l_{\pi} = 1$ \cmdthen \cmdgoto $6$ \cmdelse $j_{\pi} \gets 1$
    \State \tab $x_{\pi} \gets \FAS{Q[j_{\pi}], \bot}$
    \Statex \tab \cmdif $x_{\pi} = \bot$ \cmdthen  
    \Statex \tab[2] \cmdif $j_{\pi} = l_{\pi} - 1$ \cmdthen \cmdgoto $6$
    \Statex \tab[2] \cmdelse \{$j_{\pi} \gets j_{\pi} + 1$; \cmdgoto $8$\}
    \State \tab \cmdreturn $x_{\pi}$
\end{algorithmic}
\end{multicols}
    \vspace{-0.1in}
    \caption{
        Herlihy-Wing queue implementation \cite{HW90}.
        Each numbered line in the implementation has at most one shared memory instruction, and is performed atomically.
    }
\label{impl:queue}
\end{minipage}}\end{figure}

\subsection{The proof process}

To show that an implementation $\Ocal$ is linearizable via the tracking technique, the prover must present a tracker $\Oaug$ and a machine-certified proof that the statement $\I_L \equiv (\M \ne \nullset)$ is an invariant of $\A(\Oaug)$.
Generally, we prove $\I_L$'s invariance by induction over the length of the run.
$\I_L$ holds in the initial configuration by the tracker definition, so the base case is straightforward.
$\I_L$'s validity in subsequent configurations however, relies not only on its validity in the current configuration, but also on the design of the algorithm, i.e., other invariants of the algorithm that capture the states of the various program variables (objects and registers).
Thus, in order to go through with the induction, we must strengthen $\I_L$ to a stronger invariant $\I$ that meets two conditions:  
(a) $\I$ is {\em inductive} and (b) $\I$ implies $\I_L$.
In general, we accomplish (b) by ensuring that $\I_L$ is a conjunct in the statement $\I$ (i.e., $\I \equiv \I_L \wedge \ldots$).
The task of strengthening $\I_L$ to an inductive $\I$ is the main intellectual work that the prover must do.
The identification of $\I$ requires an understanding of ``why the algorithm works'', and thus the {\em prover} ({\em unlike the verifier}) must still understand the algorithm well in order to give the proof.
Of course, the prover must subsequently {\em prove} $\I$ by induction; but our experience through proving several algorithms suggests that this latter step, while potentially time-taking due to the length of the proof, is intellectually easy once the correct $\I$ is identified.
In summary, to prove an implementation $\Ocal$ is linearizable, the human-prover:
\begin{enumerate}
\item
Presents a tracker $\Oaug$ of the implementation $\Ocal$.
\item
Identifies an inductive invariant $\I$ that contains $\I_L$ as a conjunct.
\item
Publishes a machine-certified inductive proof that $\I$ is an invariant of $\A(\Oaug)$.
\end{enumerate}

To demonstrate the proof process, we now present our proof of the Herlihy-Wing queue.
Since the principal purpose here is to explain the proof process, rather than the queue implementation, we will focus the discussion on the process and make as few references to the particulars of the implementation as possible.
We now describe each of the three steps:

\begin{figure}[]\fbox{\begin{minipage}\textwidth		
\begin{algorithmic}[1]
    \medskip
    \Statex {\bf Base Objects:}
    \begin{itemize}[noitemsep,topsep=0pt,parsep=0pt,partopsep=0pt]
        \item $X$ is a F\&Inc/read register initialized to 1.
        \item $Q[1, 2, \ldots]$ is an infinite FAS/read/write array, where each $Q[i]$ is initialized to $\bot$.
        \augment{
        \item $\M$ initialized to $\{(\sigma_0, f_0) \}$ is a meta-configuration, where $\sigma_0$ is the empty sequence, and $f_0$ maps each process $\pi \in \Pi$ to $(\bot,\bot,\bot)$.
        }
\end{itemize}
    
\end{algorithmic}

\begin{algorithmic}[1]
    \medskip 
    \State \cmdprocedure $\Oaug.\Enqueue_{\pi}(v_{\pi})$
    \augment{
    \Statex $\M \gets \{C' \mid \exists C \in \M : C' = invoke(C, \pi, \Enqueue_{\pi}(v_{\pi}))\}$
    }
    \State \tab $i_{\pi} \gets \FAInc{X}$
    \augment{
    \Statex \tab $
  	\M  \gets \left\{ C'\ \middle\vert	
  	            \begin{array}{l}
    	                \exists C \in \M, S \subseteq pending(C), \alpha \in Perm(S): \\
    	                    \tab C' = \delta^*(C, \alpha) \wedge \ \forall \pi' \in S : pc_{\pi'} \in \{3, 4\} \vee \pi' = \pi
                    \end{array}
  	\right\}
      $
    }
    
    \State \tab $Q[i_{\pi}] \gets v_{\pi}$
    \State \tab \cmdreturn $\ack$
    \augment{
    \Statex \tab $\M \gets \{C' \mid \exists C \in \M : C' = return(C, \pi, \ack)\}$
    }
    \medskip
    
    \State \cmdprocedure $\Oaug.\Dequeue_{\pi}()$
    \augment{
    \Statex $\M \gets \{C' \mid \exists C \in \M : C' = invoke(C, \pi, \Dequeue_{\pi}())\}$
    }
    \State \tab $l_{\pi} \gets X$
    \State \tab \cmdif $l_{\pi} = 1$ \cmdthen \cmdgoto $6$ \cmdelse $j_{\pi} \gets 1$
    \State \tab $x_{\pi} \gets \FAS{Q[j_{\pi}], \bot}$
    \Statex \tab \cmdif $x_{\pi} = \bot$ \cmdthen  
    \Statex \tab[2] \cmdif $j_{\pi} = l_{\pi} - 1$ \cmdthen \cmdgoto $6$
    \Statex \tab[2] \cmdelse \{$j_{\pi} \gets j_{\pi} + 1$; \cmdgoto $8$\}
    \augment{
    \Statex \tab \cmdelse $\M \gets \{C' \mid \exists C \in \M : C' = \delta^*(C, \pi)\}$
    }
    \State \tab \cmdreturn $x_{\pi}$
    \augment{
    \Statex \tab $\M \gets \{C' \mid \exists C \in \M : C' = return(C, \pi, x_{\pi})\}$
    }
\end{algorithmic}
    \caption{
        Tracker $\Oaug$ for the queue implementation $\Ocal$ presented in Figure~\ref{impl:queue}
    }
\label{impl:queue-tracker}
\end{minipage}}\end{figure}

\begin{figure}[]\fbox{\begin{minipage}\textwidth		
\caption{Invariant $\I$ of $\A(\Oaug)$, where $\Oaug$ is the implementation of the queue tracker in Figure~\ref{impl:queue-tracker}.}
\vspace{-4mm}

\begin{align*}
\I \equiv\: \I_L &\wedge \I_{X} \wedge \I_Q \wedge \I_v \wedge \I_i \wedge \I_l \wedge \I_j \wedge \I_x \wedge \I_{pc} \wedge \I_{\M} \\
            &\wedge \I_{1,5} \wedge \I_2 \wedge \I_3 \wedge \I_4 \wedge \I_6 \wedge I_7 \wedge \I_8 \wedge \I_9 \wedge \I_a \wedge \I_b
\end{align*}

\footnotesize{In the above expression, the various conjuncts on the right hand side are defined below. Appendix~\ref{sec:queue} contains a full explanation of the invariant and the predicates used in it.}

\scriptsize{
\begin{itemize}[parsep=0pt,partopsep=0pt]
    \item
    $\I_{L} \equiv \M \ne \nullset$
    \item
    $\I_{X} \equiv X \in \N^+$
    , 
    $\I_{Q} \equiv \forall k \in \N^+: Q[k] \in \N^+ \cup \{\bot\}$
    , 
    $\I_{v} \equiv \forall \pi \in \Pi: v_\pi \in \N^+$
    , 
    $\I_{i} \equiv \forall \pi \in \Pi: i_\pi \in \N^+$
    \item[]
    $\I_{l} \equiv \forall \pi \in \Pi: l_\pi \in \N^+$
    , 
    $\I_{j} \equiv \forall \pi \in \Pi: j_\pi \in \N^+$
    , 
    $\I_{x} \equiv \forall \pi \in \Pi: x_\pi \in \N^+ \cup \{\bot\}$
    , 
    $\I_{pc} \equiv \forall \pi \in \Pi: pc_\pi \in [9]$
    \item
    $\I_\M \equiv \M \subseteq \{(\sigma, f) : \sigma \in \bigcup_{n \in N} (\N^+)^n, f \in (\{\Enqueue, \Dequeue\} \times (\N^+ \cup \{\bot\}) \times (\N^+ \cup \{\ack, \bot\}))^{\Pi}\}$
    \item
    $\I_{1, 5} \equiv \forall \pi \in \Pi : \forall (\sigma, f) \in \M : pc_{\pi} \in \{1, 5\} \implies f(\pi) = (\bot, \bot, \bot)$
    \item 
    $\I_{2} \equiv \forall \pi \in \Pi : \forall (\sigma, f) \in \M : pc_{\pi} = 2 \implies f(\pi) = (\Enqueue, v_{\pi}, \bot)$
    \item 
    $\I_{3} \equiv \forall \pi \in \Pi : \forall (\sigma, f) \in \M : pc_{\pi} = 3 \implies f(\pi) \in \{\Enqueue\} \times \{v_{\pi}\} \times \{\ack, \bot\} \wedge (1 \leq i_{\pi} < X) \wedge (Q[i_{\pi}] = \bot)$ \\
    $\hphantom{\I_{3} \equiv \forall \pi \in \Pi : \forall (\sigma, f) \in \M : pc_{\pi} = 3 \implies} \ \wedge 
    (\forall \pi' \in \Pi - \{\pi\}: pc_{\pi'} \in \{3, 4\} \implies i_{\pi'} \neq i_{\pi})$
    \item 
    $\I_{4} \equiv \forall \pi \in \Pi : \forall (\sigma, f) \in \M : pc_{\pi} = 4 \implies f(\pi) \in \{\Enqueue\} \times \{v_{\pi}\} \times \{\ack, \bot\} \wedge (1 \leq i_{\pi} < X)$ \\
    $\hphantom{\I_{4} \equiv \forall \pi \in \Pi : \forall (\sigma, f) \in \M : pc_{\pi} = 4 \implies} \ \wedge 
    (\forall \pi' \in \Pi - \{\pi\}: pc_{\pi'} \in \{3, 4\} \implies i_{\pi'} \neq i_{\pi})$
    \item 
    $\I_{6} \equiv \forall \pi \in \Pi : \forall (\sigma, f) \in \M : pc_{\pi} = 6 \implies f(\pi) = (\Dequeue, \bot, \bot)$
    \item 
    $\I_{7} \equiv \forall \pi \in \Pi : \forall (\sigma, f) \in \M : pc_{\pi} = 7 \implies f(\pi) = (\Dequeue, \bot, \bot) \wedge 1 \leq l_{\pi} \leq X$
    \item 
    $\I_{8} \equiv \forall \pi \in \Pi : \forall (\sigma, f) \in \M : pc_{\pi} = 8 \implies f(\pi) = (\Dequeue, \bot, \bot) \wedge 1 \leq j_{\pi} < l_{\pi} \leq X$
    \item
    $\I_{9} \equiv \forall \pi \in \Pi : \forall (\sigma, f) \in \M : pc_{\pi} = 9 \implies f(\pi) \in \{\Dequeue\} \times \{\bot\} \times (\N^+ \cup \{\bot\})$
    
    \item
    $\I_a \equiv \forall k \in \N^+ : k > X - 1 \implies Q[k] = \bot$ 
    \item 
    $\I_b \equiv \forall A \subseteq [X-1] :
    GoodEnqSet(A) \implies (\forall s \in Perm(A) : (JInvSeq(S) $
    \\
    $\hphantom{\I_b \equiv \forall A \subseteq [X-1] : \ ( \forall s \in Perm(A) : JInvSeq} \implies (\exists C \in \M : GoodRes(A, C) \wedge ValuesMatchInds(s, C))))$
    
\end{itemize}
}
\label{inv:queue}
\end{minipage}}\end{figure}

\begin{enumerate}
\item
\underline{Tracker}:
We present our partial tracker $\Oaug$ in Figure~\ref{impl:queue-tracker}.
To make the correspondence with the original implementation clear, we color the augmenting code \augment{\augColor}.
A simple inspection of the augmentations shows that $\Oaug$ conforms to the initialization rule, and invocation, intermediate, and return line rules of a partial tracker.

Specifying a {\em partial} tracker, as opposed to a {\em full} tracker, allows us to express our insights into exactly when and how the implemented object linearizes operations.
For instance, we have embedded our understanding that every operation can be linearized at line 2 or line 8 of some process's code by only augmenting those two intermediate lines.
Furthermore, our augmentation implies that it suffices to only track linearizations that linearize an operation by $\pi'$ when $\pi$ executes line 2 if the program counter of $\pi'$ is at line 3 or 4.
Embedding such insights into the tracker makes identifying and proving the inductive invariant $\I$ easier.

\item
\underline{Inductive invariant}:
We identify an inductive invariant $\I$ of $\Oaug$, and display it in Figure~\ref{inv:queue}.
(Appendix~\ref{sec:queue} contains a full explanation of the invariant and the predicates used in it.)
Notice that $\I_L$ is one of the conjuncts in $\I$.
The exact statement of the invariant comes from a careful study of the algorithm to determine which truths it maintains in order to be linearizable in the manner captured by the partial tracker.
A birdseye view of the invariant is as follows:
invariant conjuncts $\I_x$, where $x$ is a variable in the implementation, express truths about what values those variables take;
invariant conjuncts $\I_\ell$, where $\ell$ is a line number, express truths that can be inferred when a process $\pi$ has its program counter at line $\ell$.

\item
\underline{Machine-certified proof}:
With the partial tracker $\Oaug$ and the corresponding invariant $\I$ in hand, the actual induction proof is quite mechanical, making it a perfect fit to be checked and certified by a machine.
The proof of the induction step has eight cases, one for each line $\ell$ of the implementation, and it comprehensively justifies why each of the invariant conjuncts holds after an arbitrary process $\pi$ whose program counter currently points to $\ell$ executes that line of code.
Our inductive proof of $\I$ and the additional (trivial) claim that $\I_L \equiv (\M \ne \nullset)$ is an invariant of $\Oaug$ have been checked by the Temporal Language of Action Proof System (TLAPS).
We have made this machine-certified proof available at: {\footnotesize \url{https://github.com/uguryavuz/machine-certified-linearizability}}.
\end{enumerate}

\subsection{The verification process}

Convincing the reader that the implementation is linearizable, is very simple:
the reader simply needs to check that the augmented implementation in Figure~\ref{impl:queue-tracker} is indeed a partial tracker of the queue,
and then confirm that TLAPS has already certified that $|\M| \ne \nullset$ is an invariant of the partial tracker at: {\footnotesize \url{https://github.com/uguryavuz/machine-certified-linearizability}}.
Done!
In particular, the reader {\em does not} need to read or understand the proof of the invariant; in fact, the reader {\em does not} even need to understand the code of the implementation!
It suffices to simply observe that the machine has verified the invariant $\M \ne \nullset$.
In summary, to verify an implementation $\Ocal$ is linearizable, the human-verifier simply:
\begin{enumerate}
    \item 
    Checks that the augmentation $\Oaug$ presented by the prover is indeed a tracker of $\Ocal$.
    \item
    Confirms that a machine proof-assistant has certified that $\M \ne \nullset$ is an invariant of $\A(\Oaug)$.
\end{enumerate}

\section{Conclusion and Remarks}

We have presented the {\em tracking technique} for proving linearizability and strong linearizability of concurrent data structures, and have demonstrated the technique's efficacy by penning machine-verified proofs of algorithms that are noted for their complexity and speed. 
We look forward to many more algorithms being successfully machine-verified, both by us and by others in the community.

Due to the inherent complexity of concurrent algorithms, we believe that machine-verification can play an even wider role in providing robust, trusted guarantees.
We are extending our technique to incorporate other variants of linearizability, such as {\em durable linearizability} \cite{Izraelevitz, Attiya} and {\em strict linearizability} \cite{StrictLinearizability}.
We are also designing techniques to verify properties of mutual exclusion locks, such as starvation-freedom and first-come-first-served.
Finally, we are developing techniques to produce machine-verified proofs of time complexity guarantees of multiprocess algorithms.
We believe an interesting related open problem is the development of proof methods for consistency criteria that are weaker than linearizability, such as {\em sequential consistency}.

\bibliographystyle{acm}

\appendix
\section{Proof of the Key Lemma for Full Trackers}
\label{sec:proof-of-full-tracker-main-lemma}

\mainlemma*

\begin{proof}
		The proof is by induction on the number of events in the run $\Rfull$.
		\begin{itemize}
			\item 
			{\bf Base Case:} 
			Every implementation run $\Rfull$ with zero events is simply an initial configuration $\Cfull_0$ of algorithm $\A(\Ofull)$.
			In this case, $behavior(\Rfull)$ is the empty behavior, and thus only the empty atomic run is in $Linearizations(\Rfull) = AtomicRuns(behavior(\Rfull))$.
			The final (and only) configuration of this run is the unique initial atomic configuration $C_0 = (\sigma_0, f_0)$;
			thus $\C(Linearizations(\Rfull)) = \{(\sigma_0, f_0) \}$ which is the definition of $\Mfull(\Cfull_0)$.
			
			\item
			{\bf Induction Step:}
			Let $\Rfull_k = \Cfull_0, \ldots, \Cfull_{k-1},(\pi_k, \ell_k),\Cfull_k$ be run with $k \ge 1$ events.
			We consider the prefix-run $\Rfull_{k-1} = \Cfull_0,\ldots,\Cfull_{k-1}$, and note that the induction hypothesis states that:
			\[\Mfull(\Rfull_{k-1}) = \C(Linearizations(\Rfull_{k-1}))\]
			
			For notational convenience, we define $B_k = behavior(\Rfull_k)$ and $B_{k-1} = behavior(\Rfull_{k-1})$.
			The last line executed in $\Rfull$, i.e. $\ell_k$, can be one of three types of lines: invocation, intermediate, or return. 
			\begin{itemize}
				\item 
				\underline{Case: $\ell_k$ is an invocation $op_{\pi}(arg)$}.
				First, we show that RHS $\subseteq$ LHS.
				That is, letting $L_k \in Linearizations(\Rfull_k)$, we must first prove that $\C(L_k) \in \Mfull(\Rfull_k)$. 
				To this effect, define $L_{k-1}$ to be the longest prefix of $L_k$, such that $behavior(L_{k-1}) = B_{k-1}$.
				By definition of atomic runs, $L_k$ results when $L_{k-1}$ is followed immediately by the invocation $op_{\pi}(arg)$ and subsequently by a (possibly empty) sequence of linearization steps by processes $\alpha = \alpha_1,\ldots,\alpha_h$ that are yet to linearize in $invoke(\C(L_{k-1}), \pi, op_\pi(arg))$.
				In particular, no other invocation or return events are possible, since $behavior(L_k) = B_k$ which has just the single invocation more than $B_{k-1}$.
				Thus, by definition $\C(L_k)$ results from $\pi$ invoking $op_\pi(arg)$ and $\alpha_1,\ldots,\alpha_h$ linearizing after configuration $\C(L_{k-1})$.
				Since $\C(L_{k-1}) \in \Mfull(\Rfull_{k-1})$ by the inductive hypothesis, $\C(L_k) \in EvolveInv(\Mfull(\Rfull_{k-1}),  op_\pi(arg)) = \Mfull(\Rfull_k)$.
				
                Second we prove that LHS $\subseteq$ RHS.
                That is, let $C_k \in \Mfull(\Rfull_k)$, we must prove that there is a linearization $L_k$ of $\Rfull_k$ with final configuration $\C(L_k) = C_k$.
                To this effect, we note that $\Mfull(\Rfull_k) = EvolveInv(\Mfull(\Rfull_{k-1}), op_\pi(arg))$.
                That is, there is a configuration $C_{k-1} \in \Mfull(\Rfull_{k-1})$ such that $C_k$ results from $\pi$ invoking $op_\pi(arg)$ and some sequence of processes $\alpha = \alpha_1,\ldots,\alpha_h$ linearizing after $C_{k-1}$.
                By the inductive hypothesis, there is a linearization $L_{k-1}$ of $\Rfull_{k-1}$ whose final configuration is $\C(L_{k-1}) = C_{k-1}$ in which $\pi$ is idle and $\alpha_1,\ldots,\alpha_h$ are yet-to-linearize in $invoke(C_{k-1}, \pi, op_\pi(arg))$.
                Thus, the atomic run $L_k$ that extends $L_{k-1}$ with $\pi$ invoking and $\alpha$ linearizing after $C_{k-1}$ is a linearization of $\Rfull_k$ whose final configuration is $\C(L_k) = C_k$, which concludes the proof of the case.
                
				\item
				\underline{Case: $\ell_k$ is an intermediate line}.
                Since $behavior(\Rfull_k) = behavior(\Rfull_{k-1})$,
                we invoke the inductive hypothesis to get the equality:
                \[\Mfull(\Rfull_{k-1}) = \C(Linearizations(\Rfull_{k-1})) = \C(Linearizations(\Rfull_k)) \]
                Now, we observe that when $\Mfull(\Rfull_{k-1})$ is already the full set of final configurations of linearizations with the behavior $B_{k-1} = B_{k}$, evolving does not change the set, that is:
                \[\Mfull(\Rfull_k) = Evolve(\Mfull(\Rfull_{k-1})) = Evolve(\C(Linearizations(\Rfull_k))) = \C(Linearizations(\Rfull_k))\]
                Thus, we conclude the proof of the case.
                                
				\item
				\underline{Case: $\ell_k$ is a return $\text{\bf return } res$}.
				This case is very similar to the first case.
				Once again, we first show that RHS $\subseteq$ LHS.
				That is, let $L_k \in Linearizations(\Rfull_k)$; we must first prove that $\C(L_k) \in \Mfull(\Rfull_k)$. 
				To this effect, define $L_{k-1}$ to be the longest prefix of $L_k$, such that $behavior(L_{k-1}) = B_{k-1}$.
				By definition of atomic runs, $L_k$ results when $L_{k-1}$ is followed immediately by the return of $res$ and subsequently by a (possibly empty) sequence of linearization steps by processes $\alpha = \alpha_1,\ldots,\alpha_h$ that are yet to linearize in $return(\C(L_{k-1}), \pi,  res)$.
				In particular, no other invocation or return events are possible, since $behavior(L_k) = B_k$ which has just the single return more than $B_{k-1}$.
				Thus, by definition $\C(L_k)$ results from $\pi$ returning $res$ and $\alpha_1,\ldots,\alpha_h$ linearizing after configuration $\C(L_{k-1})$.
				Since $\C(L_{k-1}) \in \Mfull(\Rfull_{k-1})$ by the inductive hypothesis, $\C(L_k) \in EvolveRet(\Mfull(\Rfull_{k-1}), \pi, res) = \Mfull(\Rfull)$.
				
                Second we prove that LHS $\subseteq$ RHS.
                That is, let $C_k \in \Mfull(\Rfull_k)$, we must prove that there is a linearization $L_k$ of $\Rfull_k$ with final configuration $\C(L_k) = C_k$.
                To this effect, we note that $\Mfull(\Rfull_k) = EvolveRet(\Mfull(\Rfull_{k-1}), \pi, res)$.
                That is, there is a configuration $C_{k-1} \in \Mfull(\Rfull_{k-1})$ such that $C_k$ results from $\pi$ returning $res$ and some sequence of processes $\alpha = \alpha_1,\ldots,\alpha_h$ linearizing after $C_{k-1}$.
                By the inductive hypothesis, there is a linearization $L_{k-1}$ of $\Rfull_{k-1}$ whose final configuration is $\C(L_{k-1}) = C_{k-1}$ in which $\pi$ has linearized with return value $res$ and $\alpha_1,\ldots,\alpha_h$ are yet-to-linearize in $return(C_{k-1}, \pi, res)$.
                Thus, the atomic run $L_k$ that extends $L_{k-1}$ with $\pi$ returning and $\alpha$ linearizing after $C_{k-1}$ is a linearization of $\Rfull_k$ whose final configuration is $\C(L_k) = C_k$, which concludes the proof of the case.
			\end{itemize}			 
		\end{itemize}	
\end{proof}

\section{Proof of Strong Linearizability Verification Theorem via Partial Trackers}
\label{sec:strong-linearizability-theorem}

\thmstronglinearizability*

\begin{proof}
    We split the proof into two parts, proving the {\em only if} direction, and then proving the {\em if} direction.
    \begin{enumerate}
    \item 
    To prove the {\em only if} direction, we assume there is a linearization function $\Lcal$ that maps each finite run $R_k = C_0,(\pi_1, \ell_1),C_1,\ldots,(\pi_k, \ell_k),C_k$ of $\A(\Ocal)$ to a linearization $\Lcal(R_k)$.
    We claim there is a tracker $\Opart$ which maintains a meta-configuration $\Mpart$, whose value $\Mpart(\Rpart_k)$ after any run $\Rpart_k$ that is coupled with run $R_k$ of $\Ocal$ is equal to $\Mpart(\Rpart_k) = \{\Lcal(R_k)\}$.
    We prove this claim by induction.
    \begin{itemize}
    \item[]
    {\em Base Case:}
    note that if $R_k = R_0$ is a zero-event run, then its only linearization is the zero-event atomic run $L_0 = C_0$ that (starts and) ends in the initial atomic configuration $(\sigma_0, f_0)$, thus $\Lcal(R_0) = L_0$ and $\Mpart$'s initialization is indeed $\{(\sigma_0, f_0)\}$, which concludes the base case.
    
    \item[]
    {\em Induction Step:}
    if $\Mpart(\Rpart_k) = \{\Lcal(R_k) \}$ for some $k \ge 0$, and the run $R_k$ extends to $R_{k+1}$ in a single step, then by assumption $L_{k+1} = \Lcal(R_{k+1})$ is an extension of $L_k = \Lcal(R_k)$.
    Here, we break the argument into cases, depending on which type of line $\ell_{k+1}$ is (invocation, intermediate, or return):
    \begin{enumerate}
	 	\item
    	\underline{Case: $\ell$ is the invocation of $op_{\pi}(arg)$}.
    	$\C(L_{k+1})$ must result from some processes $\alpha_1,\ldots,\alpha_{h_1}$ linearizing, then $\pi_{k+1}$ invoking $op_\pi(arg)$, then $\alpha_{h_1 + 1},\ldots,\alpha_{h_2}$ linearizing after $\C(L_k)$ for some $0 \le h_1 \le h_2$.
    	However, since linearizing an operation before or after an invocation makes no difference to the final configuration,
    	we can obtain the same $\C(L_{k+1})$ after $\C(L_k)$ by $\pi$ invoking $op_\pi(arg)$ and then linearizing $\alpha_1,\ldots,\alpha_{h_2}$.
    	Thus, $\C(L_{k+1}) \in EvolveInv(\Mpart(\Rpart_k), op_\pi(arg))$.
    
    	\item
    	\underline{Case: $\ell$ is an intermediate line of $op_{\pi}(arg)$}.
        $\C(L_{k+1})$ must result from linearizing some sequence of processes $\alpha_1,\ldots,\alpha_h$ after $\C(L_k)$.
    	Thus, $\C(L_{k+1}) \in Evolve(\Mpart(\Rpart_k))$.
        
    	\item
    	\underline{Case: $\ell$ is a $\cmdreturn \: res$ statement from a procedure $op_{\pi}(arg)$}.
    	$\C(L_{k+1})$ must result from some processes $\alpha_1,\ldots,\alpha_{h_1}$ linearizing, then $\pi_{k+1}$ returning $res$ then $\alpha_{h_1 + 1},\ldots,\alpha_{h_2}$ linearizing after $\C(L_k)$ for some $0 \le h_1 \le h_2$.
    	However, since linearizing an operation before or after a return makes no difference to the final configuration,
    	we can obtain the same $\C(L_{k+1})$ after $\C(L_k)$ by $\pi$ returning $res$ and then linearizing $\alpha_1,\ldots,\alpha_{h_2}$.
    	Thus, $\C(L_{k+1}) \in EvolveRet(\Mpart(\Rpart_k), \pi, res)$.
    \end{enumerate}
    
    \end{itemize}
    That completes the proof of the {\em only if} direction.
    
    \item
    To prove the {\em if} direction, we assume that there is a tracker $\Opart$ that always maintains a singleton meta-configuration $\Mpart$.
    We now recursively define a prefix preserving linearization function $\Lcal$ on runs of $\A(\Ocal)$, such that for coupled runs $R$ of $\Ocal$ and $\Rpart$ of $\Opart$, $\Mpart(\Rpart) = \C(\Lcal(R))$:
    \begin{itemize}
        \item 
        If $R_0$ is a zero event run of $\A(\Ocal)$, define $\Lcal(R_0) \triangleq (\sigma_0, f_0)$.
        (This is clearly a (indeed the only) linearization of $R_0$.)
        \item
        If $R_k$ is a $k$ event run for $k \ge 0$, consider the prefix run $R_{k-1}$ such that $R_k = R_{k-1},(\pi_k, \ell_k),\C(R_k)$. 
        Also consider the corresponding $k-1$ and $k$ event coupled runs of $\Opart$: $\Rpart_{k-1}$ and $\Rpart_k$, and their meta-configurations: $\Mpart(\Rpart_{k-1}) = \{C_{k-1}\}$ and $\Mpart(\Rpart_k) = \{C_k\}$.
        We now finish the definition with three cases:
        \begin{enumerate}
    	 	\item
        	\underline{Case: $\ell$ is the invocation of $op_{\pi}(arg)$}.
            $C_k$ must result from $\pi$ invoking $op_\pi(arg)$ and some sequence of processes $\alpha_1,\ldots,\alpha_h$ linearizing after $C_{k-1} = \Lcal(R_{k-1})$.
            We define $\Lcal(R_k)$ to be the run resultant from $\Lcal(R_{k-1})$ being extended by that invocation and those linearization events.
            
        	\item
        	\underline{Case: $\ell$ is an intermediate line of $op_{\pi}(arg)$}.
            $C_k$ must result from some sequence of processes $\alpha_1,\ldots,\alpha_h$ linearizing after $C_{k-1} = \Lcal(R_{k-1})$.
            We define $\Lcal(R_k)$ to be the run resultant from $\Lcal(R_{k-1})$ being extended by those linearization events.
            
        	\item
        	\underline{Case: $\ell$ is a $\cmdreturn \: res$ statement from a procedure $op_{\pi}(arg)$}.
            $C_k$ must result from $\pi$ returning $res$ and some sequence of processes $\alpha_1,\ldots,\alpha_h$ linearizing after $C_{k-1} = \Lcal(R_{k-1})$.
            We define $\Lcal(R_k)$ to be the run resultant from $\Lcal(R_{k-1})$ being extended by that return and those linearization events.
        \end{enumerate}
    \end{itemize}
    By construction, we see that $\Lcal$ is a prefix preserving linearization function of the runs of implementation $\Ocal$.
    That concludes the proof.
    \end{enumerate}
\end{proof}

\section{Herlihy-Wing Queue Full Invariant}
\label{sec:queue}

We present the strengthened invariant $\I$ of the Herlihy-Wing queue implementation in Figure~\ref{inv:queue-old}. 
$\I$ is a conjunction, with $\I_L$ among its conjuncts.
Thus, when we prove its invariance, we have $\I_L$ by implication.
The remaining conjuncts can be understood as follows:
\begin{itemize}
    \item The conjuncts $\I_X, \I_Q, \I_v, \I_i, \I_l, \I_j, \I_x, \I_{pc}, \I_{\M}$ express type safety. 
    That is, the various variables in the algorithm always take on values that we would expect satisfy their types. 
    \item The conjuncts $\I_{1, 5}, \I_2, \I_3, \I_4, \I_6, \I_7, \I_8, \I_9$ express truths that pertain to a process $\pi$ when its program counter $pc_{\pi}$ has a particular value, both regarding the meta-configurations permitted by the suggested tracker, as well as the values of the variables of the original implementation.
    \item The conjuncts $\I_a$ and $\I_b$ express general truths about the implementation and the tracker. $\I_a$ asserts that for any index $k$ of $Q$ greater than $X - 1$, it is the case that $Q[k] = \bot$, which should be trivial since $X$ denotes the foremost unused index of $Q$. The discussion of $\I_b$ requires defining a number of additional predicates first.
\end{itemize}
\begin{figure}[!htb]\fbox{\begin{minipage}\textwidth		
\caption{Invariant $\I$ of $\A(\Oaug)$, where $\Oaug$ is the implementation of the queue tracker in Figure~\ref{impl:queue-tracker}.}
\vspace{-4mm}
\begin{align*}
\I \equiv\: &\I_L  \\
            &\wedge\: \I_{X} \wedge \I_Q \wedge \I_v \wedge \I_i \wedge \I_l \wedge \I_j \wedge \I_x \wedge \I_{pc} \wedge \I_{\M} \\
            &\wedge\: \I_{1,5} \wedge \I_2 \wedge \I_3 \wedge \I_4 \wedge \I_6 \wedge I_7 \wedge \I_8 \wedge \I_9 \\
            &\wedge\: \I_a \wedge \I_b
\end{align*}

In the above expression, the various conjuncts on the right hand side are defined below.

\begin{itemize}[parsep=0pt,partopsep=0pt]
    \item
    $\I_{L} \equiv \M \ne \nullset$
    \item
    $\I_{X} \equiv X \in \N^+$
    \item
    $\I_{Q} \equiv \forall k \in \N^+: Q[k] \in \N^+ \cup \{\bot\}$
    \item
    $\I_{v} \equiv \forall \pi \in \Pi: v_\pi \in \N^+$
    \item
    $\I_{i} \equiv \forall \pi \in \Pi: i_\pi \in \N^+$
    \item
    $\I_{l} \equiv \forall \pi \in \Pi: l_\pi \in \N^+$
    \item
    $\I_{j} \equiv \forall \pi \in \Pi: j_\pi \in \N^+$
    \item
    $\I_{x} \equiv \forall \pi \in \Pi: x_\pi \in \N^+ \cup \{\bot\}$
    \item
    $\I_{pc} \equiv \forall \pi \in \Pi: pc_\pi \in [9]$
    \item
    $\I_\M \equiv \M \subseteq \{(\sigma, f) : \sigma \in \bigcup_{n \in N} (\N^+)^n,$
    \\
    $\hphantom{\I_\M \equiv \M \subseteq \{(\sigma, f) : \ } f \in (\{\Enqueue, \Dequeue\} \times (\N^+ \cup \{\bot\}) \times (\N^+ \cup \{\ack, \bot\}))^{\Pi}\}$
    \item
    $\I_{1, 5} \equiv \forall \pi \in \Pi : \forall (\sigma, f) \in \M : pc_{\pi} \in \{1, 5\} \implies f(\pi) = (\bot, \bot, \bot)$
    \item 
    $\I_{2} \equiv \forall \pi \in \Pi : \forall (\sigma, f) \in \M : pc_{\pi} = 2 \implies f(\pi) = (\Enqueue, v_{\pi}, \bot)$
    \item 
    $\I_{3} \equiv \forall \pi \in \Pi : \forall (\sigma, f) \in \M : pc_{\pi} = 3 \implies f(\pi) \in \{\Enqueue\} \times \{v_{\pi}\} \times \{\ack, \bot\}$ \\
    $\hphantom{\I_{3} \equiv \forall \pi \in \Pi : \forall (\sigma, f) \in \M : pc_{\pi} = 3 \implies} \ \wedge (1 \leq i_{\pi} < X) \wedge (Q[i_{\pi}] = \bot)$\\
    $\hphantom{\I_{3} \equiv \forall \pi \in \Pi : \forall (\sigma, f) \in \M : pc_{\pi} = 3 \implies} \ \wedge 
    (\forall \pi' \in \Pi - \{\pi\}: pc_{\pi'} \in \{3, 4\} \implies i_{\pi'} \neq i_{\pi})$
    \item 
    $\I_{4} \equiv \forall \pi \in \Pi : \forall (\sigma, f) \in \M : pc_{\pi} = 4 \implies f(\pi) \in \{\Enqueue\} \times \{v_{\pi}\} \times \{\ack, \bot\}$ \\
    $\hphantom{\I_{4} \equiv \forall \pi \in \Pi : \forall (\sigma, f) \in \M : pc_{\pi} = 4 \implies} \ \wedge (1 \leq i_{\pi} < X)$\\
    $\hphantom{\I_{4} \equiv \forall \pi \in \Pi : \forall (\sigma, f) \in \M : pc_{\pi} = 4 \implies} \ \wedge 
    (\forall \pi' \in \Pi - \{\pi\}: pc_{\pi'} \in \{3, 4\} \implies i_{\pi'} \neq i_{\pi})$
    \item 
    $\I_{6} \equiv \forall \pi \in \Pi : \forall (\sigma, f) \in \M : pc_{\pi} = 6 \implies f(\pi) = (\Dequeue, \bot, \bot)$
    \item 
    $\I_{7} \equiv \forall \pi \in \Pi : \forall (\sigma, f) \in \M : pc_{\pi} = 7 \implies f(\pi) = (\Dequeue, \bot, \bot) \wedge 1 \leq l_{\pi} \leq X$
    \item 
    $\I_{8} \equiv \forall \pi \in \Pi : \forall (\sigma, f) \in \M : pc_{\pi} = 8 \implies f(\pi) = (\Dequeue, \bot, \bot) \wedge 1 \leq j_{\pi} < l_{\pi} \leq X$
    \item
    $\I_{9} \equiv \forall \pi \in \Pi : \forall (\sigma, f) \in \M : pc_{\pi} = 9 \implies f(\pi) \in \{\Dequeue\} \times \{\bot\} \times (\N^+ \cup \{\bot\})$
    
    \item
    $\I_a \equiv \forall k \in \N^+ : k > X - 1 \implies Q[k] = \bot$ 
    \item 
    $\I_b \equiv \forall A \subseteq [X-1] :
    GoodEnqSet(A)$
    \\
    $\hphantom{\I_b \equiv \forall A \subseteq [X-1] : \ } \implies (\forall s \in Perm(A) : 
    (JInvSeq(s) $ \\
    $\hphantom{\I_b \equiv \forall A \subseteq [X-1] : \ ( \forall s} \implies (\exists C \in \M : GoodRes(A, C) \wedge ValuesMatchInds(s, C))))$
    
 \end{itemize}
\label{inv:queue-old}
\end{minipage}}\end{figure}

Let us first define a predicate $GoodEnqSet$ that, given a set $A$ of indices between $1$ and $X-1$, 
determines whether the set can correspond to a set of linearized $\Enqueue$ operations.
If an index $k \in [X-1]$ corresponds to a non-$\bot$ element stored in $Q$, then it comes from a linearized $\Enqueue$ operation and must appear in $A$.
Otherwise, an index $k$ that appears in $A$ and is such that $Q[k] = \bot$, must appear in $A$ by virtue of the existence of a process $\pi$ at line 3, that has claimed $k$ as the index of $Q$ it will write its argument into.
We can then formally define $GoodEnqSet(A)$ for any $A \subseteq [X-1]$ as follows:
\begin{align*}
GoodEnqSet(A) \triangleq \forall k \in [X-1] : \ & (Q[k] \neq \bot \implies k \in A) \\ &\wedge ((Q[k] = \bot \wedge k \in A) \implies \exists \pi' \in \Pi : pc_{\pi'} = 3 \wedge i_{\pi'} = k).
\end{align*}

Let us then consider the responses processes should receive under specific linearizations. If a process $\pi$ is at line 4 or 9, its operation must have been linearized and the process must have been assigned a return value of either $\ack$ or $x_{\pi}$, respectively for $\Enqueue$ and $\Dequeue$. Moreover, it must have been assigned a response value of $\ack$ at line 3 if $\pi$, or some other process, linearized $\pi$'s $\Enqueue$ as part of their update to $\M$ in line 2. With this observation, let $A$ be a conjectured set of the indices of $Q$ between $1$ and $X-1$, claimed for use by the linearized $\Enqueue$ operations.
Then, any $\pi$ at line 3 whose picked index $i_{\pi}$ appears in $A$, must also be linearized and also assigned a return value of $\ack$. Besides these three cases, no other processes are assigned return values.
In the light of these observations, for a conjectured set of linearized $\Enqueue$ indices $A \subseteq [X-1]$ and a configuration $C \in \M$, let us define the predicate $GoodRes(A, C)$ which indicates that the response field of $C.f$ has return values that correspond to the linearizations conjectured by $A$:
\begin{align*}
    GoodRes(A, C) \triangleq \forall \pi' \in \Pi: C.f(\pi').res = \begin{cases}
        \ack     & \text{if } pc_{\pi'} = 3 \wedge i_{\pi'} \in A \\
        \ack     & \text{if } pc_{\pi'} = 4 \\
        x_{\pi'} & \text{if } pc_{\pi'} = 9 \\
        \bot     & \text{otherwise.}
    \end{cases}
\end{align*}

Akin to how we define a valid correspondence between the response fields of configurations in $\M$ and return values resulting from particular conjectured linearizations in $GoodRes$, we can also define a correspondence between the state fields of configurations and the values of variables in the implementation.
If $s$, a permutation of $A \subseteq [X-1]$, is a sequence of indices in the order which the corresponding enqueue operations are conjectured to linearize,
then the state of the implemented object should be a sequence of values pointed to by the indices (or in the case where the value pointed to by the index is $\bot$, the argument of the process having picked the given index), in the order the indices appear in $s$. 
Then, for any $s \in Perm(A)$ where $A \subseteq [X-1]$ and a configuration $C \in \M$, let us define the predicate $ValuesMatchInds(s, C)$, which indicates that $C.\sigma$ indeed corresponds to the order of linearizations conjectured by the sequence of indices $s$:
\begin{align*}
    ValuesMatchInds(s, C) \triangleq
    C.\sigma = (\alpha_1, \alpha_2, \ldots, \alpha_{|s|}) \text { where } \\ \alpha_k = 
    \begin{cases}
        Q[s_k] & \text{if } Q[s_k] \neq \bot \\
        v_{\pi'} & \text{where } \pi' \in \Pi : pc_{\pi'} = 3 \wedge i_{\pi'} = s_{k}.
    \end{cases}
\end{align*}

We note that this is a well-defined expression for all $A$ where $GoodEnqSet(A)$ holds true, as this would (along with invariant $\I_3$) guarantee the existence of a unique process $\pi'$ where the second case is applicable.

For the final necessary definition, let us once again consider a permutation $s$ of a set $A \subseteq [X-1]$ of the conjectured set of indices claimed for use by the linearized $\Enqueue$ operations, and reflect about what might render $s$ a potentially correct sequence of linearizations.
Before all else, if $s$ is simply the sequence of non-empty indices of $Q$ in the order in which they appear in $Q$, it should clearly represent a potentially correct sequence of linearizations, since an incoming dequeuing process can execute successive $\Dequeue$ operations and indeed empty the queue in this order, meaning that this sequence corresponds to a potentially correct state of the implemented object.
Let us consider the cases where $s$ does not follow this order, and suppose there exists a pair $m, n$ such that $n < m$ yet $s_m < s_n$.
In this case, if $s_m$ points to an empty component in $Q$,
$s$ constitutes a potentially correct ordering of the indices, as a dequeuing process might very well pass through $s_m$ as it loops through indices, before reaching $s_n$.
However, if $Q[s_m] \neq \bot$, it is not immediately clear whether $s$ is a potentially correct ordering.
In order to be able to justify this, we must ensure that there is a dequeuing process $\pi$ which is past $s_m$ in its loop (i.e. $s_m < j_{\pi}$), such that $s_n < l_{\pi}$ (which would otherwise fall outside the scope of indices $\pi$ could dequeue from).

Then, to capture this idea, let an \textit{inversion} for a sequence $s$ of indices of $Q$, denote a pair $m, n \in [|s|]$ for which $n < m, s_m < s_n$ and $Q[s_m] \neq \bot$.
For a sequence of indices to correspond to a potentially correct sequence of linearizations, the inversions it contains must all be \textit{justified}. A \textit{justified inversion}, as discussed in the preceding paragraph, is an inversion for which there is a dequeuing process $\pi$ at line 8, such that $s_n < l_{\pi}$ and $s_m < j_{\pi}$.
We can formally define this as a predicate $JInvSeq(s)$, for $s \in Perm(A)$ for some $A \subseteq [X-1]$, as follows:
\begin{align*}
    JInvSeq(s) \triangleq
    \forall m, n \in [|s|] : (n < m &\wedge s_m < s_n \wedge Q[s_m] \neq \bot) \\ &\implies (\exists \pi' \in \Pi : pc_{\pi'} = 8 \wedge s_n < l_{\pi'} \wedge s_m < j_{\pi'}).
\end{align*}

This finishes the definition of $\I$. 
Our TLAPS-certified proof of this invariant and the linearizability of the Herlihy-Wing queue can be found at: 
{\footnotesize
\url{https://github.com/uguryavuz/machine-certified-linearizability}}.

\clearpage
\section{The Jayanti-Tarjan Union-Find Object}
\label{sec:union-find}

In this section, we consider Jayanti and Tarjan's concurrent union-find implementation \cite{JT16, JayantiTarjanBoix, JT21}, and describe our TLAPS certified proof of its strong linearizability.
We chose Jayanti and Tarjan's algorithm due its extensive use in practice---it is the fastest algorithm for computing connected components of a graph on CPUs \cite{dhulipala2020connectit} and GPUs \cite{GpuUnionFind}, and has several other applications \cite{alistarh2019search}.

\subsection{The Union-Find Type}

The union-find type maintains a partition of the {\em elements} in $[n] = \{1,2,\ldots,n\}$ and supports two operations.
\begin{itemize}
    \item 
    $\Find(x)$ returns the maximum element in element $x$'s part of the partition.
    \item 
    $\Unite(x,y)$ merges the parts containing $x$ and $y$ if they are different and returns $\ack$.
\end{itemize}
Formally, we specify the partition as a function $\sigma:[n] \to \mathcal{P}([n])$ from the set of elements to the powerset of the set of elements, such that $\sigma(x)$ is the part containing element $x$.
Of course, if two elements $x$ and $y$ are in the same part of the partition, then $\sigma(x) = \sigma(y)$.
We give the full formal specification of the union-find type in the figure Object Type~\ref{type:union-find}.

\begin{figure}[h]\fbox{\begin{minipage}\textwidth		
\begin{type}[Union-Find Object]
A union-find type of $n$ elements $[n] = \{1,\ldots,n\}$ is described as follows:
\label{type:union-find}
\begin{itemize}[parsep=0pt,partopsep=0pt]
    \item 
    $\Sigma = \Bigl\{\sigma: [n] \to \mathcal{P}([n]) \Bigm| \{\sigma(x) \mid x \in [n]\} \text{ is a partition of } [n] \text{, and } \forall x \in [n]: x \in \sigma(x) \Bigr\}$
    \item
    $OP = \{\Unite{}, \Find{}\}$
    \item 
    $ARG_{\Unite{}} = [n] \times [n], ARG_{\Find{}} = [n]$
    \item
    $RES = \{\ack\} \cup [n]$
    \item
    Transition function $\delta$ is defined by:
    \begin{itemize}
        \item 
        $\delta(\sigma, \pi, \Find{}, x) = \max \sigma(x)$
        \item
            $\delta(\sigma, \pi, \Unite{}, (x, y)) =
            \begin{cases}
              (\sigma, \ack),  & \text{if}\ \sigma(x) = \sigma(y) \\
              (\sigma', \ack), & \text{if}\ \sigma(x) \ne \sigma(y) \\
                               & \text{where}\ \forall z \notin \sigma(x) \cup \sigma(y), \sigma'(z) = \sigma(z)   \\
                               & \text{and}\ \hspace{0.13in} \forall z \in \sigma(x) \cup \sigma(y), \sigma'(z) = \sigma(x) \cup \sigma(y)
            \end{cases}$
    \end{itemize}
\end{itemize}
\end{type}
\end{minipage}}\end{figure}

\subsection{The Jayanti-Tarjan Union-Find Implementation}

We present Jayanti and Tarjan's implementation of union-find in Figure~\ref{impl:splitting-uf}.
Each numbered line in the implementation requires the performance of at most one shared memory instruction, and is performed atomically.

\begin{figure}[!htb]\fbox{\begin{minipage}\textwidth		
    \begin{algorithmic}[1]
        \Statex
        \Statex {\bf Base Objects:}  
				\begin{itemize}
				\item
    		$x.par$ is a Read/CAS register initialized to $x.par = x$, for each node $x \in [n]$.
				\end{itemize}
    \end{algorithmic}
    
    \begin{algorithmic}[1]	
        \State \cmdprocedure $\Ocal.\Find_\pi(x_\pi)$
        \Statex \tab $u_\pi \gets x_\pi$ 
        \State \tab $a_{\pi} \gets u_{\pi}.par$
        \Statex \tab \cmdif $u_\pi = a_\pi$ \cmdthen 
        \Statex \tab[2] \cmdgoto~line~6
        \State \tab $b_\pi \gets a_{\pi}.par$
        \State \tab $\CAS{u_\pi.par, a_\pi, b_\pi}$; \cmdgoto~line~2~or~5
        \State \tab $u_\pi \gets a_\pi$; \cmdgoto~line~2
        \State  \tab \cmdreturn $u_\pi$

        \Statex

        \State  \cmdprocedure $\Ocal.\Unite_\pi(x_\pi, y_\pi)$ 
        \Statex \tab $u_\pi \gets x_\pi$; $v_\pi \gets y_\pi$
        \State  \tab \cmdif $u_\pi = v_\pi$ \cmdthen \cmdgoto~line~17
        \Statex \tab \cmdelif $u_\pi < v_\pi$ \cmdthen \cmdif $\CAS{u_\pi.par, u_\pi, v_\pi}$ \cmdthen \cmdgoto~line~17
        \Statex \tab \cmdelif $u_\pi > v_\pi$ \cmdthen \cmdif $\CAS{v_\pi.par, v_\pi, u_\pi}$ \cmdthen \cmdgoto~line~17
        \State \tab $a_{\pi} \gets u_{\pi}.par$
        \Statex \tab \cmdif $u_\pi = a_\pi$ \cmdthen \cmdgoto~line~13
        \State \tab $b_\pi \gets a_{\pi}.par$
        \State \tab $\CAS{u_\pi.par, a_\pi, b_\pi}$; \cmdgoto~line~9~or~12
        \State \tab $u_\pi \gets a_\pi$; \cmdgoto~line~9
        \State \tab $a_{\pi} \gets v_{\pi}.par$
        \Statex \tab \cmdif $v_\pi = a_\pi$ \cmdthen \cmdgoto~line~8
        \State \tab $b_\pi \gets a_{\pi}.par$
        \State \tab $\CAS{v_\pi.par, a_\pi, b_\pi}$; \cmdgoto~line~13~or~16
        \State \tab $v_\pi \gets a_\pi$; \cmdgoto~line~13
        \State \tab \cmdreturn \ack
    \end{algorithmic}
    
    \caption{ 
        Jayanti and Tarjan's implementation of a union-find object on $n$ nodes $\{1,2,\ldots,n\}$ each initially in its own singleton part of the partition, i.e., the initial state is $\sigma_0: [n] \to \mathcal{P}([n])$ defined by $\forall x \in [n], \sigma_0(x) = x$.
    }
    \label{impl:splitting-uf}
\end{minipage}}\end{figure}

\paragraph{\bf A special note about our implementation}
In the implementation, each element is represented by a node, and each node $z$ has a parent pointer field $z.par$ that points to another node.
Each part of the partition is represented as a single parent pointer tree, so $\sigma(z) = \sigma(z.par)$, and 
we maintain the invariant that the parent of $z$ is always greater than or equal to $z$ (i.e. $z.par \ge z$).
Thus, the roots of the trees are the largest elements in their respective partition; if $w$ is a root of its tree than $w.par = w$.
In the initial state $\sigma_0$, all elements are in their own singleton part of the partition, i.e. $\forall z \in [n], \sigma_0(z) = \{z\}$.
Correspondingly, the implementation starts with $\forall z \in [n], z.par = z$.

With this representation, a process $\pi$ could perform $\Find_\pi(x_\pi)$ by starting a node $u_\pi$ at $x_\pi$ (line 1), walking $u_\pi$ up the parent pointers until it reaches a root, and returning that root, which must be the maximum element in the set.
However, for efficiency of future find operations, the Jayanti and Tarjan observed that it helps to {\em compact the tree}, i.e., change the parent pointers of the intermediate nodes encountered along the $x_\pi$-to-root {\em find path} to point closer to the root.
They presented two variants of compaction for their algorithm \cite{JT21}, called: {\em one-try splitting} and {\em two-try splitting}.
In one-try splitting: for each intermediate node $u_\pi$ on the find path, the implementation attempts to {\em improve} $u_\pi.par$ from its parent $a_\pi$ (line 2), to its grand-parent $b_\pi$ (line 3) via a CAS, and then moves on to the next node (the {\bf goto} line 5 path on line 4).
In two-try splitting: the implementation attempts to improve each $u_\pi.par$ again (the {\bf goto} line 2 path on line 4).
Using non-determinism at line 4, we have incorporated both variants into a single implementation.
In fact, our implementation allows any number of tries on each intermediate node (by taking the {\bf goto} line 2 path repeatedly), and varying number of tries on different nodes.
Our proof thereby applies to a wide class of concurrent union-find variants that we dub ``any-try splitting''.
Incidentally, two-try splitting has the better theoretical efficiency bound \cite{JT21}, but one-try splitting seems to perform slightly better in practice on most test cases \cite{dhulipala2020connectit, GpuUnionFind}.
To our knowledge, other variations of any-try splitting are yet to be tested in practice.
Splitting is also done when the implementation walks up the path from $x_\pi$ and $y_\pi$ in the $\Unite$ procedure (lines 9-16), and we similarly apply our non-determinism strategy to incorporate any-try splitting (lines 11 and 15).
Otherwise, our implementation mimics the original implementation of Jayanti and Tarjan \cite{JT16}.

\clearpage
\subsection{The Tracker}

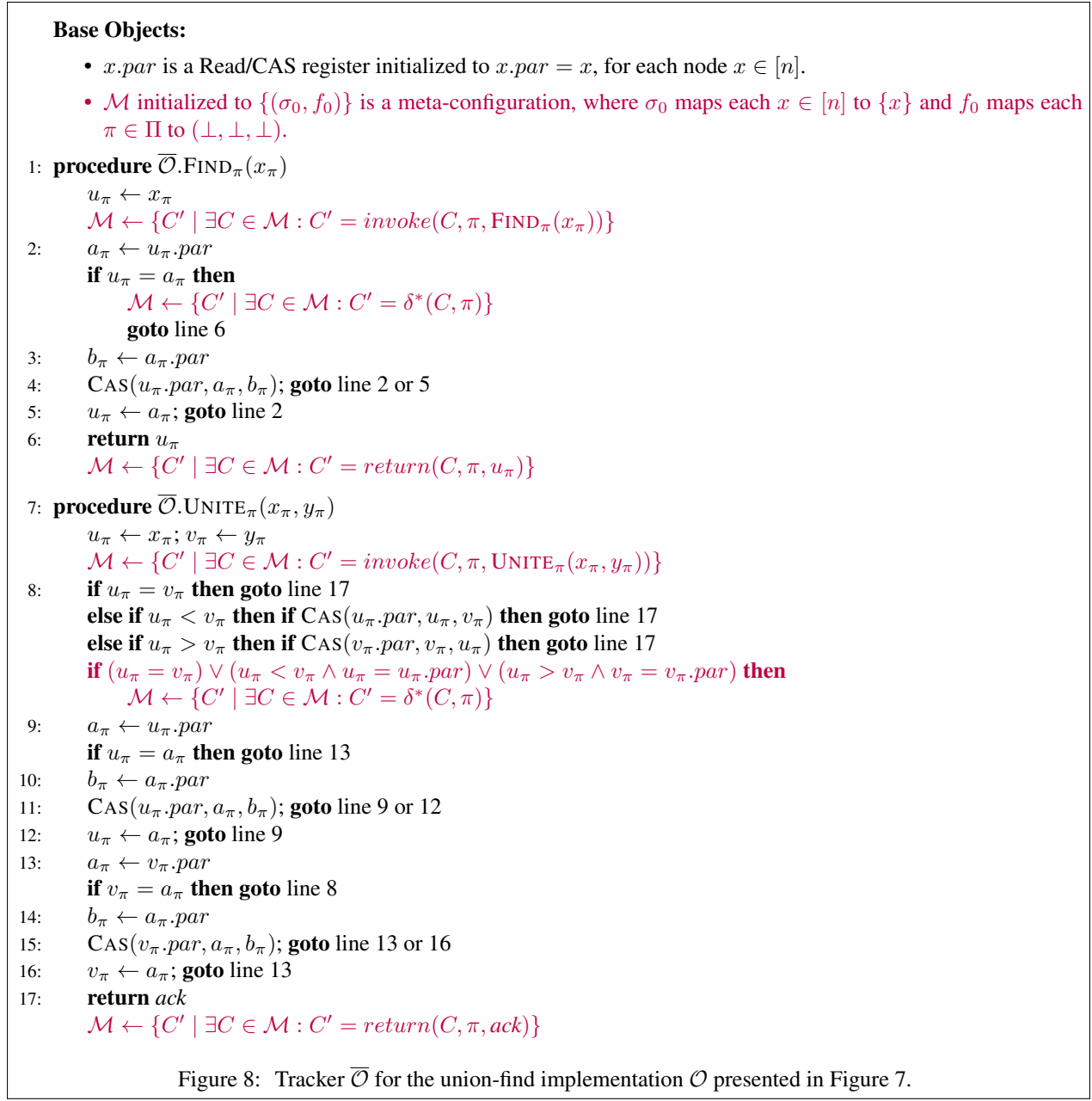
\begin{figure}[H]\fbox{\begin{minipage}\textwidth		
    \begin{algorithmic}[1]
        \Statex
        \Statex {\bf Base Objects:}  
        \begin{itemize}
            \item
            $x.par$ is a Read/CAS register initialized to $x.par = x$, for each node $x \in [n]$.
            \augment{
            \item
            $\M$ initialized to $\{(\sigma_0, f_0) \}$ is a meta-configuration, where $\sigma_0$ maps each $x \in [n]$ to $\{x\}$ and $f_0$ maps each $\pi \in \Pi$ to $(\bot,\bot,\bot)$.
            }
        \end{itemize}
    \end{algorithmic}
    
    \begin{algorithmic}[1]	
        \State \cmdprocedure $\Oaug.\Find_\pi(x_\pi)$
        \Statex \tab $u_\pi \gets x_\pi$ 
        \augment{
        \Statex \tab 
        $\M \gets \left\{ C' \mid	
                \exists C \in \M:
                    C' = invoke(C, \pi, \Find_\pi(x_\pi))
        \right\}$
        }
        \State \tab $a_{\pi} \gets u_{\pi}.par$
        \Statex \tab \cmdif $u_\pi = a_\pi$ \cmdthen 
        \augment{
        \Statex \tab[2]
        $\M \gets \left\{ C' \mid
                    \exists C \in \M:
                        C' = \delta^*(C, \pi)
        \right\}$
        }
        \Statex \tab[2] \cmdgoto~line~6
        \State \tab $b_\pi \gets a_{\pi}.par$
        \State \tab $\CAS{u_\pi.par, a_\pi, b_\pi}$; \cmdgoto~line~2~or~5
        \State \tab $u_\pi \gets a_\pi$; \cmdgoto~line~2
        \State  \tab \cmdreturn $u_\pi$
        \augment{
        \Statex \tab 
        $\M \gets \left\{ C' \mid	
                \exists C \in \M:
                    C' = return(C, \pi, u_\pi)
        \right\}$
        }
        \medskip

        \State \cmdprocedure $\Oaug.\Unite_\pi(x_\pi, y_\pi)$ 
        \Statex \tab $u_\pi \gets x_\pi$; $v_\pi \gets y_\pi$
        \augment{
        \Statex \tab 
        $\M \gets \left\{ C' \mid	
                \exists C \in \M:
                    C' = invoke(C, \pi, \Unite_\pi(x_\pi, y_\pi))
        \right\}$
        }
        \State  \tab \cmdif $u_\pi = v_\pi$ \cmdthen \cmdgoto~line~17
        \Statex \tab \cmdelif $u_\pi < v_\pi$ \cmdthen \cmdif $\CAS{u_\pi.par, u_\pi, v_\pi}$ \cmdthen \cmdgoto~line~17
        \Statex \tab \cmdelif $u_\pi > v_\pi$ \cmdthen \cmdif $\CAS{v_\pi.par, v_\pi, u_\pi}$ \cmdthen \cmdgoto~line~17
        \augment{
        \Statex \tab 
        \cmdif $(u_\pi = v_\pi) \vee (u_\pi < v_\pi \wedge u_\pi = u_\pi.par) \vee (u_\pi > v_\pi \wedge v_\pi = v_\pi.par)$ \cmdthen 
        \Statex \tab[2]   	
        $\M \gets \left\{ C' \mid	
                \exists C \in \M:
                    C' = \delta^*(C, \pi)
        \right\}$
        }
        \State \tab $a_{\pi} \gets u_{\pi}.par$
        \Statex \tab \cmdif $u_\pi = a_\pi$ \cmdthen \cmdgoto~line~13
        \State \tab $b_\pi \gets a_{\pi}.par$
        \State \tab $\CAS{u_\pi.par, a_\pi, b_\pi}$; \cmdgoto~line~9~or~12
        \State \tab $u_\pi \gets a_\pi$; \cmdgoto~line~9
        \State \tab $a_{\pi} \gets v_{\pi}.par$
        \Statex \tab \cmdif $v_\pi = a_\pi$ \cmdthen \cmdgoto~line~8
        \State \tab $b_\pi \gets a_{\pi}.par$
        \State \tab $\CAS{v_\pi.par, a_\pi, b_\pi}$; \cmdgoto~line~13~or~16
        \State \tab $v_\pi \gets a_\pi$; \cmdgoto~line~13
        \State \tab \cmdreturn \ack
        \augment{
        \Statex \tab 
        $\M \gets \left\{ C' \mid	
                \exists C \in \M:
                    C' = return(C, \pi, \ack)
        \right\}$
        }
    \end{algorithmic}
    
    \caption{ 
		Tracker $\Oaug$ for the union-find implementation $\Ocal$ presented in Figure~\ref{impl:splitting-uf}.
    }
    \label{impl:splitting-uf-tracker}
\end{minipage}}\end{figure}

We present our partial tracker in Figure~\ref{impl:splitting-uf-tracker}.
Since we are proving {\em strong linearizability}, rather than just linearizability, we choose a partial tracker that will ensure that there will be exactly one configuration in the meta-configuration variable $\M$ at any point in time.
Our key insight into the implementation to write down this tracker is our ability to identify a unique linearization point for each operation.

\clearpage
\subsection{Proving The Invariant}

Our task is to prove that the statement $\I_S \equiv (|\M| = 1)$ is an invariant of $\A(\Oaug)$ in order to deduce that $\Ocal$ is strongly linearizable.
Of course, $\I_S$ implies $\I_L \equiv (\M \ne \nullset)$, so we also obtain the result that the Jayanti-Tarjan union-find is linearizable.
(Of course, any strongly linearizable implementation is linearizable, so there is no surprise here.)

We present our strengthened invariant $\I$ in Figure~\ref{inv:splitting-uf}, which is conjunction with $\I_S$ and $\I_L$ as conjuncts.
Our TLAPS-verified proof of $\I$, and the corollaries that $\I_S$ and $\I_L$ are invariants of $\Oaug$ are publicly available in the GitHub repository, and certify that all variants of the Jayanti-Tarjan union-find object are strongly linearizable.

\begin{figure}[H]\fbox{\begin{minipage}\textwidth		
\caption{
Invariant $\I$ of $\A(\Oaug)$, where $\Oaug$ is the union-find tracker in Figure~\ref{impl:splitting-uf-tracker}.
}
\begin{align*}
\I \equiv\: &\I_L \wedge \I_S \\
            &\wedge\: \I_{par} \wedge \I_x \wedge \I_y \wedge \I_u \wedge \I_v \wedge \I_a \wedge \I_b \wedge \I_{pc} \wedge \I_\M \\
            &\wedge\: \I_{UF1} \wedge \I_{UF2} \wedge \I_{UF3} \wedge \I_{UF4} \wedge \I_{UF5} \\
            &\wedge\: \I_{1,7} \wedge \I_2 \wedge \I_3 \wedge \I_{4, 5} \wedge \I_6 \wedge \I_{8, 9, 13} \wedge \I_{10} \wedge \I_{11, 12} \wedge \I_{14} \wedge \I_{15, 16} \wedge \I_{17} 
\end{align*}

In the above expression, the various conjuncts on the right hand side are defined below.

\begin{footnotesize}
\begin{itemize}[parsep=0pt,partopsep=0pt]
    \item
    $\I_{L} \equiv \M \ne \nullset$
    \item 
    $\I_{S} \equiv |\M| = 1$
    \item
    $\I_{par} \equiv \forall z \in [n]: z.par \in [n]$
    \item
    $\I_{x} \equiv \forall \pi \in \Pi: x_\pi \in [n]$
    \item
    $\I_{y} \equiv \forall \pi \in \Pi: y_\pi \in [n]$
    \item
    $\I_{u} \equiv \forall \pi \in \Pi: u_\pi \in [n]$
    \item
    $\I_{v} \equiv \forall \pi \in \Pi: v_\pi \in [n]$
    \item
    $\I_{a} \equiv \forall \pi \in \Pi: a_\pi \in [n]$
    \item
    $\I_{b} \equiv \forall \pi \in \Pi: b_\pi \in [n]$
    \item
    $\I_{pc} \equiv \forall \pi \in \Pi: pc_\pi \in [17]$
    \item
    $\I_\M \equiv \M \subseteq \{(\sigma, f) : \sigma \in \mathcal{P}([n])^{[n]}, f \in (\{\Find, \Unite\} \times ([n] \cup [n]^2]) \times ([n] \cup \{\ack, \bot\}))^{\Pi}\}$
    \item
    $\I_{UF1} \equiv \forall z \in [n]: z.par \ge z$
    \item
    $\I_{UF2} \equiv \forall z \in [n] : \forall (\sigma, f) \in \M : z \in \sigma(z)$
    \item 
    $\I_{UF3} \equiv \forall w, z \in [n] : \forall (\sigma, f) \in \M : (w \in \sigma(z) \implies \sigma(w) = \sigma(z))$ 
    \item
    $\I_{UF4} \equiv \forall w, z \in [n] : \forall (\sigma, f) \in \M : (w.par = z \implies \sigma(w) = \sigma(z))$ 
    \item 
    $\I_{UF5} \equiv \forall w, z \in [n] : \forall (\sigma, f) \in \M : ((w \neq z \wedge w.par = w \wedge z.par = z) \implies \sigma(w) \neq \sigma(z))$ 
    \item
    $\I_{1, 7} \equiv 
        \forall \pi \in \Pi : \forall (\sigma, f) \in \M : 
        pc_\pi \in \{1, 7\} \implies 
        f(\pi) = (\bot, \bot, \bot)$
    \item
    $\I_2 \equiv 
        \forall \pi \in \Pi : \forall (\sigma, f) \in \M : 
        pc_\pi = 2 \implies 
        (\sigma(u_\pi) = \sigma(x_\pi) \wedge 
        f(\pi) = (\Find, x_\pi, \bot))$
    \item
    $\I_3 \equiv 
        \forall \pi \in \Pi : (\forall (\sigma, f) \in \M : 
        pc_\pi = 3 \implies 
        (\sigma(u_\pi) = \sigma(x_\pi) = \sigma(a_\pi) \wedge 
        f(\pi) = (\Find, x_\pi, \bot))) \wedge a_\pi \geq u_\pi$
    \item
    $\I_{4, 5} \equiv 
        \forall \pi \in \Pi : (\forall (\sigma, f) \in \M : 
        pc_\pi \in \{4, 5\} \implies 
        (\sigma(u_\pi) = \sigma(x_\pi) = \sigma(a_\pi) = \sigma(b_\pi)$ \\ 
    $\hphantom{\I_{4, 5} \equiv 
    \forall \pi \in \Pi : (\forall (\sigma, f) \in \M : 
    pc_\pi \in \{4, 5\} \implies \ \ } \wedge 
        f(\pi) = (\Find, x_\pi, \bot))) \wedge b_\pi \geq a_\pi \geq u_\pi$
    \item
    $\I_6 \equiv 
        \forall \pi \in \Pi : \forall (\sigma, f) \in \M : 
        pc_\pi = 6 \implies 
        f(\pi) = (\Find, x_\pi, u_\pi))$
    \item
    $\I_{8, 9, 13} \equiv 
        \forall \pi \in \Pi : \forall (\sigma, f) \in \M : 
        pc_\pi \in \{8, 9, 13\} \implies 
        (\sigma(u_\pi) = \sigma(x_\pi) \wedge 
        \sigma(v_\pi) = \sigma(y_\pi)$ \\
    $\hphantom{\I_{8, 9, 13} \equiv 
    \forall \pi \in \Pi : \forall (\sigma, f) \in \M : 
    pc_\pi \in \{8, 9, 13\} \implies \ \ }
        \wedge f(\pi) = (\Unite, (x_\pi, y_\pi), \bot))$
    \item
    $\I_{10} \equiv
        \forall \pi \in \Pi : (\forall (\sigma, f) \in \M : 
        pc_\pi = 10 \implies 
        (\sigma(u_\pi) = \sigma(x_\pi) = \sigma(a_\pi)) \wedge 
        \sigma(v_\pi) = \sigma(y_\pi)$ \\
    $\hphantom{\I_{10} \equiv
    \forall \pi \in \Pi : (\forall (\sigma, f) \in \M : 
    pc_\pi = 10 \implies \ \ } 
        \wedge f(\pi) = (\Unite, (x_\pi, y_\pi), \bot))) \wedge
        a_\pi \geq u_\pi$
    \item
    $\I_{11, 12} \equiv
        \forall \pi \in \Pi : (\forall (\sigma, f) \in \M : 
        pc_\pi \in \{11, 12\} \implies 
        (\sigma(u_\pi) = \sigma(x_\pi) = \sigma(a_\pi) = \sigma(b_\pi)) $ \\ 
    $\hphantom{\I_{11, 12} \equiv
    \forall \pi \in \Pi : (\forall (\sigma, f) \in \M : 
    pc_\pi \in \{11, 12\} \implies ( } 
    \wedge \sigma(v_\pi) = \sigma(y_\pi)$ \\ 
    $\hphantom{\I_{11, 12} \equiv
    \forall \pi \in \Pi : (\forall (\sigma, f) \in \M : 
    pc_\pi \in \{11, 12\} \implies ( } 
        \wedge f(\pi) = (\Unite, (x_\pi, y_\pi), \bot))) \wedge
        b_\pi \geq a_\pi \geq u_\pi$
    \item
    $\I_{14} \equiv
        \forall \pi \in \Pi : (\forall (\sigma, f) \in \M : 
        pc_\pi = 14 \implies 
        (\sigma(v_\pi) = \sigma(y_\pi) = \sigma(a_\pi) \wedge \sigma(u_\pi) = \sigma(x_\pi))$ \\ 
    $\hphantom{\I_{14} \equiv
    \forall \pi \in \Pi : (\forall (\sigma, f) \in \M : 
    pc_\pi = 14 \implies ( } 
        \wedge f(\pi) = (\Unite, (x_\pi, y_\pi), \bot))) \wedge
        a_\pi \geq v_\pi$
    \item
    $\I_{15, 16} \equiv
        \forall \pi \in \Pi : (\forall (\sigma, f) \in \M : 
        pc_\pi \in \{15, 16\} \implies 
        (\sigma(v_\pi) = \sigma(y_\pi) = \sigma(a_\pi)) = \sigma(b_\pi)) $ \\ 
    $\hphantom{\I_{15, 16} \equiv
    \forall \pi \in \Pi : (\forall (\sigma, f) \in \M : 
    pc_\pi \in \{15, 16\} \implies ( }
        \wedge \sigma(u_\pi) = \sigma(x_\pi) $ \\ 
    $\hphantom{\I_{15, 16} \equiv
    \forall \pi \in \Pi : (\forall (\sigma, f) \in \M : 
    pc_\pi \in \{15, 16\} \implies ( }
        \wedge f(\pi) = (\Unite, (x_\pi, y_\pi), \bot))) \wedge
        b_\pi \geq a_\pi \geq v_\pi$
    \item
    $\I_{17} \equiv 
        \forall \pi \in \Pi : \forall (\sigma, f) \in \M : 
        pc_\pi = 17 \implies 
        f(\pi) = (\Unite, (x_\pi, y_\pi), \ack))$
 \end{itemize}
 \end{footnotesize}
\label{inv:splitting-uf}

\end{minipage}}\end{figure}

\newpage
\section{Jayanti's Single-Writer Single-Scanner Snapshot}
\label{sec:snapshot}

In this example, we apply our technique to Jayanti's single-scanner, single-writer snapshot object implementation \cite{PJayanti05} and produce a machine-certified proof of its linearizability by TLAPS.
Below, we simply present the formal data type definition, the implementation and our partial tracker for it, and the complete inductive invariant.
Recall once again, that in order to verify the linearizability of the implementation, it suffices to check that the partial tracker is indeed well-formed and confirm that TLAPS has verified the invariant $\M \ne \nullset$.

\subsection{The Single-Scanner, Single-Writer Snapshot Type}

An $m$-component snapshot object embodies an array of $m$ elements from $\N^+$ that supports two operations:
\begin{itemize}
    \item $\Write(i, v)$ writes $v$ into the $i$th component of the array.
    \item $\Scan()$ retrieves the
state of the array.
\end{itemize}
Additionally, a single-scanner, single-writer snapshot object has two constraints: (1) no two $\Scan()$ operations may be concurrent with each other, and (2) no two $\Write(i, v)$ and $\Write(i, v')$ operations may be concurrent with each other. 
Therefore, for any implementation $\Ocal$ of the single-scanner, single-writer snapshot, the generator algorithm $\A(\Ocal)$ picks operations and arguments such that the following property is maintained:
\begin{align*}
    \I_{SWSS} \equiv \forall \pi, \pi' \in \Pi : op_{\pi} = op_{\pi'} \wedge \pi \neq \pi' \implies op_{\pi} = \Write \land i_{\pi} \neq i_{\pi'}
\end{align*}

We provide the full formal specification of the single-writer, single-scanner snapshot object in the figure Object Type~\ref{type:snapshot}.
\begin{figure}[h]\fbox{\begin{minipage}\textwidth		
\begin{type}[Single-writer, single-scanner snapshot]
A single-writer, single-scanner snapshot of size $m$ with elements from $\N^+$ is described as follows:
\label{type:snapshot}
\begin{itemize}[parsep=0pt,partopsep=0pt]
    \item 
    $\Sigma = (\N^+)^m$ 
    \item
    $OP = \{\Write, \Scan \}$
    \item
    $ARG_{\Write} = (\{0, \ldots, m - 1\},\N^+), ARG_{\Scan} = \{\bot\}$.
    \item
    $RES = \{\ack \} \cup (\N^+)^m$ 
    \item
    Transition function $\delta$ is defined as follows:
    \begin{itemize}
        \item $\delta(\sigma, \pi, \Write, (i, v)) = (\sigma', \ack)$, where $\sigma'[i] = v \wedge \forall k \in \{0, \ldots, m - 1\} : k \neq i \implies \sigma[i] = \sigma'[i]$
        \item $\delta(\sigma, \pi, \Scan, \bot) = (\sigma, \sigma)$
    \end{itemize}
\end{itemize}
\end{type}
\end{minipage}}\end{figure}

\newpage
\subsection{The Jayanti Single-Writer, Single-Scanner Snapshot Implementation}

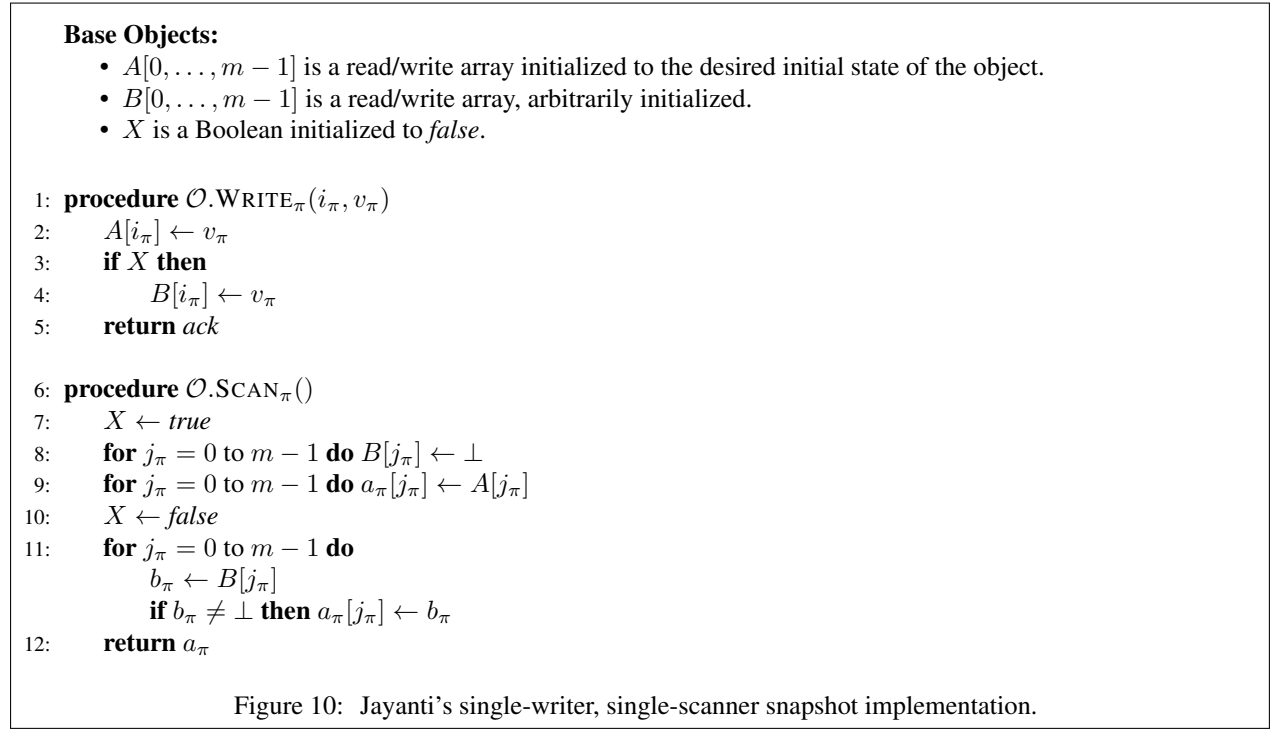
\begin{figure}[!htb]\fbox{\begin{minipage}\textwidth		
\begin{algorithmic}[1]
    \Statex
    \Statex {\bf Base Objects:}
    \begin{itemize}[noitemsep,topsep=0pt,parsep=0pt,partopsep=0pt]
        \item $A[0, \ldots, m - 1]$ is a read/write array initialized to the desired initial state of the object.
        \item $B[0, \ldots, m - 1]$ is a read/write array, arbitrarily initialized.
        \item $X$ is a Boolean initialized to $\false$.
    \end{itemize}
\end{algorithmic}

\begin{algorithmic}[1]
    \Statex 
    \State \cmdprocedure $\Ocal.\Write_{\pi}(i_{\pi}, v_{\pi})$
    \State \tab $A[i_{\pi}] \gets v_{\pi}$
    \State \tab \cmdif $X$ \cmdthen
    \State \tab[2] $B[i_{\pi}] \gets v_{\pi}$
    \State \tab \cmdreturn $\ack$
    \Statex
    \State \cmdprocedure $\Ocal.\Scan_{\pi}()$
    \State \tab $X \gets \true$
    \State \tab \cmdfor $j_{\pi} = 0$ to $m - 1$ \cmddo $B[j_{\pi}] \gets \bot$
    \State \tab \cmdfor $j_{\pi} = 0$ to $m - 1$ \cmddo $a_{\pi}[j_{\pi}] \gets A[j_{\pi}]$
    \State \tab $X \gets \false$
    \State \tab \cmdfor $j_{\pi} = 0$ to $m - 1$ \cmddo
    \Statex \tab[2] $b_{\pi} \gets B[j_{\pi}]$
    \Statex \tab[2] \cmdif $b_{\pi} \neq \bot$ \cmdthen $a_{\pi}[j_{\pi}] \gets b_{\pi}$
    \State \tab \cmdreturn $a_{\pi}$
\end{algorithmic}
    \caption{
        Jayanti's single-writer, single-scanner snapshot implementation.
    }
\label{impl:snapshot}
\end{minipage}}\end{figure}

We present the Jayanti implementation of the single-scanner, single-writer object in Figure \ref{impl:snapshot}. 
As before, each numbered line in the implementation requires the performance of at most one shared memory instruction and is performed atomically. For brevity, we omit the modified generator $\A(\Ocal)$ and the constraint $\I_{SWSS}$.

\newpage
\subsection{The Tracker}

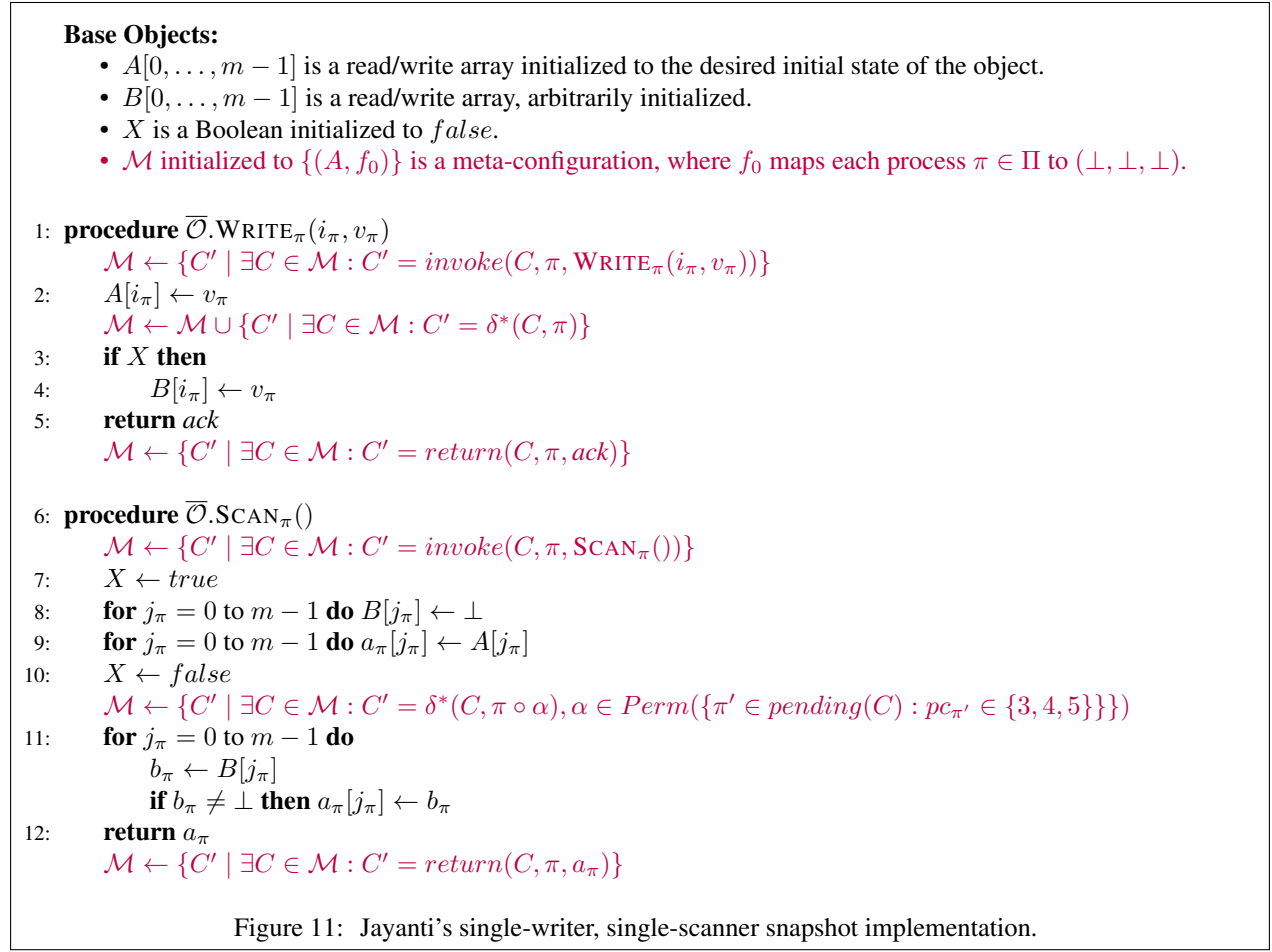
\begin{figure}[!htb]\fbox{\begin{minipage}\textwidth		
\begin{algorithmic}[1]
    \Statex
    \Statex {\bf Base Objects:}
    \begin{itemize}[noitemsep,topsep=0pt,parsep=0pt,partopsep=0pt]
        \item $A[0, \ldots, m - 1]$ is a read/write array initialized to the desired initial state of the object.
        \item $B[0, \ldots, m - 1]$ is a read/write array, arbitrarily initialized.
        \item $X$ is a Boolean initialized to $false$.
        \augment{
        \item $\M$ initialized to $\{(A, f_0) \}$ is a meta-configuration, where $f_0$ maps each process $\pi \in \Pi$ to $(\bot,\bot,\bot)$.
        }
    \end{itemize}
\end{algorithmic}

\begin{algorithmic}[1]
    \Statex 
    \State \cmdprocedure $\Oaug.\Write_{\pi}(i_{\pi}, v_{\pi})$
    \augment{
    \Statex \tab $\M \gets \{C' \mid \exists C \in \M : C' = invoke(C, \pi, \Write_{\pi}(i_{\pi}, v_{\pi}))\}$
    }
    \State \tab $A[i_{\pi}] \gets v_{\pi}$
    \augment{
    \Statex \tab $\M \gets \M \cup \{C' \mid \exists C \in \M : C' = \delta^*(C, \pi)\}$
    }
    \State \tab \cmdif $X$ \cmdthen
    \State \tab[2] $B[i_{\pi}] \gets v_{\pi}$
    \State \tab \cmdreturn $\ack$
    \augment{
    \Statex \tab $\M \gets \{C' \mid \exists C \in \M : C' = return(C, \pi, \ack)\}$
    }
    \Statex
    \State \cmdprocedure $\Oaug.\Scan_{\pi}()$
    \augment{
    \Statex \tab $\M \gets \{C' \mid \exists C \in \M : C' = invoke(C, \pi, \Scan_{\pi}())\}$
    }
    \State \tab $X \gets true$
    \State \tab \cmdfor $j_{\pi} = 0$ to $m - 1$ \cmddo $B[j_{\pi}] \gets \bot$
    \State \tab \cmdfor $j_{\pi} = 0$ to $m - 1$ \cmddo $a_{\pi}[j_{\pi}] \gets A[j_{\pi}]$
    \State \tab $X \gets false$
    \augment{
    \Statex \tab $\M \gets \{C' \mid \exists C \in \M : C' = \delta^*(C, \pi \circ \alpha), \alpha \in Perm(\{\pi' \in pending(C) : pc_{\pi'} \in \{3, 4, 5\}\} \})$
    }
    \State \tab \cmdfor $j_{\pi} = 0$ to $m - 1$ \cmddo
    \Statex \tab[2] $b_{\pi} \gets B[j_{\pi}]$
    \Statex \tab[2] \cmdif $b_{\pi} \neq \bot$ \cmdthen $a_{\pi}[j_{\pi}] \gets b_{\pi}$
    \State \tab \cmdreturn $a_{\pi}$
    \augment{
    \Statex \tab $\M \gets \{C' \mid \exists C \in \M : C' = return(C, \pi, a_{\pi})\}$
    }
\end{algorithmic}
    \caption{
        Jayanti's single-writer, single-scanner snapshot implementation.
    }
\label{impl:snapshot-tracker}
\end{minipage}}\end{figure}

Our partial tracker of the implementation 
is shown in Figure \ref{impl:snapshot-tracker}. We leverage our knowledge that this implementation linearizes operations at lines 2 and 10 by augmenting these two lines.

\newpage
\subsection{Proving The Invariant}

To prove that the statement $I_L \equiv \M \neq \nullset$ is an invariant of $\A(\Oaug)$, where $\Oaug$ is the implementation of the snapshot tracker, we begin with a few technical definitions that will help us build our inductive invariant $I$. First, let $s$ be the unique element in $S \triangleq \{\pi \in \Pi : pc_{\pi} \in \{7, \ldots, 12\}\}$ or $\bot$ if $S$ is empty. This clearly denotes the unique active scanning process, if there exists one.

Let us then consider the possible return values for $a_s$ for $s \neq \bot$ and, in particular, the set of values that could be returned for the $k$th element of the snapshot by $\Scan_s$. More specifically, for $k \in \{0, \ldots, m - 1\}$, we define $KthReturnSet(k)$ to be the set of valid values for $\sigma[k]$ before $s$ executes line 10 and that of $C.f(s).res[k]$ after $s$ executes line 10 in any tracked configuration $(\sigma, f) \in \M$ and specify its value as follows:

\begin{itemize}
    \item If $s = \bot$ or $pc_{s} = 7$, $KthReturnSet(k) = \{A[k]\}$.
    \item If $pc_{s} = 8$, \begin{align*}
        KthReturnSet(k) = \begin{cases}
        \{A[k], B[k]\} & \text{ if } k < j_{s} \wedge B[k] \neq \bot\\
        \{A[k]\} & \text{ otherwise }
    \end{cases}
    \end{align*}
    \item If $pc_{s} = 9$,
    \begin{align*}
        KthReturnSet(k) = \begin{cases}
        \{A[k], B[k]\} & \text{if } B[k] \neq \bot\\
        \{A[k], a_{\pi}[k]\} & \text{if } k < j_{s} \wedge B[k] \neq \bot\\
        \{A[k]\} & \text{ otherwise }
    \end{cases}
    \end{align*}
    \item If $pc_{s} = 10$,
    \begin{align*}
        KthReturnSet(k) = \begin{cases}
        \{A[k], B[k]\} & \text{if } B[k] \neq \bot\\
        \{A[k], a_{s}[k]\} & \text{otherwise}\\
    \end{cases}
    \end{align*}
    \item If $pc_{s} = 11$,
    \begin{align*}
        KthReturnSet(k) = \begin{cases}
                 \{A[k], B[k]\} &WB(k) \wedge B[k] \neq \bot \wedge k \geq j_s\\
                 \{A[k], a_s[k]\}&WB(k) \wedge B[k] = \bot \wedge k \geq j_s\\
                 \{B[k]\} &\neg WB(k) \wedge B[k] \neq \bot \wedge k \geq j_s\\
                 \{a_s[k]\} &\text{otherwise}
    \end{cases}
    \end{align*}
    where $WB(k) \triangleq \exists \pi \in \Pi : pc_{\pi} = 4 \wedge i_{\pi} = k$, a predicate that essentially denotes whether there is a writer that could write to the $k$th component of $B$.
    \item If $pc_{s} = 12$, $KthReturnSet(k) = \{a_s[k]\}$.
\end{itemize}

Thus, the set of possible values for $\sigma$ before $s$ executes line 10 and that of $C.f(s).res$ after $s$ executes line 10 is $ScanReturnSet$, defined as
\begin{align*}
ScanReturnSet \triangleq KthReturnSet(0) \times \cdots \times KthReturnSet(m - 1)
\end{align*}

This definition helps us make some observations about configurations $(\sigma, f)$ that are guaranteed to be in $\M$ and to complete the inductive invariant $\I$ in Figure \ref{inv:snapshot} that proves $I_L$ is an invariant of $\Oaug$.
Our TLAPS-certified proof of the invariant and hence the linearizability of Jayanti's single-writer single-scanner snapshot object can be found at: 
{\footnotesize
\url{https://github.com/uguryavuz/machine-certified-linearizability}}.

\begin{figure}[]\fbox{\begin{minipage}\textwidth		
\caption{Invariant $\I$ of $\A(\Oaug)$, where $\Oaug$ is the implementation of the snapshot tracker in Figure~\ref{impl:snapshot-tracker}.}
\vspace{-4mm}
\begin{align*}
\I \equiv\: &\I_{L}\\
            & \wedge \I_{X} \wedge \I_A \wedge \I_B \wedge \I_{\M} \wedge \I_a \wedge \I_j \wedge \I_{i} \wedge \I_{v} \wedge \I_{pc}\\
            & \wedge \I_{1, 6} \wedge \I_{2} \wedge \I_{3, 4, 5} \wedge \I_{7, 8, 9, 10} \wedge \I_{11, 12} \\
            &\wedge \I_{C} \wedge \I_{W1} \wedge \I_{W2} \wedge \I_{W3} \wedge \I_{W4} 
            \wedge \I_{S1} \wedge \I_{S2} \wedge \I_{S3} \\
            &\wedge \I_{\M 1} \wedge \I_{\M 2}
\end{align*}

In the above expression, the various conjuncts on the right hand side are defined below.

\begin{itemize}[parsep=0pt,partopsep=0pt]
    \item $\I_{L} \equiv \M \ne \nullset$
    \item $\I_X \equiv X \in \{true, false\}$
    \item $\I_A \equiv A \in (\N^+)^m$
    \item $\I_B \equiv B \in (\N^+ \cup \{ \bot \})^m$
    \item $\I_\M \equiv \M \subseteq \{(\sigma, f) : \sigma \in (\N^+)^m, f \in (\{\Write, \Scan\} \times (\{0, \ldots, m - 1\} \times (\N^+) \cup \{\bot\})$ \\ 
    $\hphantom{\I_\M \equiv \M \subseteq \{(\sigma, f) : \sigma \in (\N^+)^m, f \in (\{\Write, \Scan\} } \times ((\N^+)^m \cup \{\ack, \bot\}))^{\Pi}\}$
    \item $I_a \equiv a \in (\N^+)^m$
    \item $\I_j \equiv \forall \pi \in \Pi : j_{\pi} \in \{0, \ldots, m\}$
    \item $\I_{i} \equiv \forall \pi \in \Pi : i_{\pi} \in \{0, \ldots, m - 1\}$
    \item $\I_{v} \equiv \forall \pi \in \Pi : v_{\pi} \in \N^+$
    \item $\I_{pc} \equiv \forall \pi \in \Pi : pc_{\pi} \in [12]$
    
    \item $\I_{1, 6} \equiv \forall \pi \in \Pi: \forall (\sigma, f) \in \M : pc_{\pi} \in \{1, 6\} \implies f(\pi) = (\bot, \bot, \bot)$
    
    \item $\I_2 \equiv \forall \pi \in \Pi: \forall (\sigma, f) \in \M : pc_{\pi} = 2 \implies \sigma[i_{\pi}] = A[i_{\pi}] \wedge f(\pi) =  (\Write, (i_{\pi}, v_{\pi}), \bot)$
    
    \item $\I_{3, 4, 5} \equiv \forall \pi \in \Pi: \forall (\sigma, f) \in \M : pc_{\pi} \in \{3, 4, 5\} 
        \implies f(\pi) \in (\Write, (i_{\pi}, v_{\pi})) \times \{\ack, \bot\}$
    
    
    \item $\I_{7, 8, 9, 10} \equiv \forall \pi \in \Pi: \forall (\sigma, f) \in \M : pc_{\pi} \in \{7, 8, 9, 10\} \implies f(\pi) = (\Scan, \bot, \bot)$

    \item $\I_{11, 12} \equiv \forall \pi \in \Pi: \forall (\sigma, f) \in \M : pc_{\pi} \in \{11, 12\} \implies f(\pi) \in (\Scan, \bot) \times ((\N^+)^m \cup \{\bot\})$
    
    \item $\I_{C} \equiv \forall \pi, \pi' \in \Pi : (pc_{\pi}, pc_{\pi'} \in \{2, \ldots, 5\} \land i_{\pi} = i_{\pi'}) \lor (pc_{\pi}, pc_{\pi'} \in \{7, \ldots, 12\}) \implies \pi = \pi'$
    
    \item $\I_{W1} \equiv \forall \pi \in \Pi: \forall (\sigma, f) \in \M : (pc_{\pi} \in \{3, 4, 5\} \land \sigma[i_{\pi}] \neq A[i_{\pi}])$\\
    $\hphantom{\I_{W1} \equiv \forall \pi \in \Pi: \forall (\sigma, f) \in \M \ \ } \implies f(\pi) = (\Write, (i_{\pi}, v_{\pi}), \bot)$
    
    \item $\I_{W2} \equiv \forall \pi \in \Pi: \forall C \in \M : pc_{\pi} \in \{3, 4, 5\} \implies \exists C' \in M: C' = \delta^*(C, \pi)$
    
    \item $\I_{W3} \equiv \forall \pi \in \Pi: pc_{\pi} \in \{3, 4, 5\} \implies A[i_{\pi}] = v_{\pi}$
    
    \item $\I_{W4} \equiv \forall k \in \{0, \ldots, m - 1\}, \forall (\sigma, f) \in \M : (\forall \pi \in \Pi : \neg(pc_{\pi} \in \{3, 4, 5\} \land i_{\pi} = k))$\\
    $\hphantom{\I_{W4} \equiv \forall k \in \{0, \ldots, m - 1\}, \forall (\sigma, f) \in \M : } \implies \sigma[k] = A[k]$
    
    \item $\I_{S1} \equiv \exists \pi \in \Pi : pc_{\pi} \in \{8, 9, 10\} \iff X = true$
    
    \item $\I_{S2} \equiv \forall \pi \in \Pi : \forall k \in \{0, \ldots, m - 1\} : ((pc_{\pi} = 10 \lor (pc_{\pi} = 9 \land k < j_{\pi})) \land B[k] = \bot$ \\
    $\hphantom{\I_a \equiv \forall \pi \in \Pi : \forall k \in \{0, \ldots, m - 1\} : (( } \land (\forall \pi' \in \Pi : \neg(i_{\pi'} = k \land pc_{\pi'} \in \{3, 4\})))$ \\ 
    $\hphantom{\I_a \equiv \forall \pi \in \Pi : \forall k \in \{0, \ldots, m - 1\} :  \land} \implies A[k] = a_{\pi}[k]$
    
    \item $\I_{S3} \equiv \forall \pi \in \Pi : \forall k \in \{0, \ldots, m - 1\} : ((pc_{\pi} \in \{9, 10\} \lor (pc_{\pi} = 8 \land k < j_{\pi})) \land B[k] \neq \bot $\\
    $\hphantom{\I_B \equiv \forall \pi \in \Pi : \forall k \in \{0, \ldots, m - 1\} : (( }\land (\forall \pi' \in \Pi : \neg(i_{\pi'} = k \land pc_{\pi'} \in \{3, 4\})))$ \\ 
    $\hphantom{\I_B \equiv \forall \pi \in \Pi : \forall k \in \{0, \ldots, m - 1\} : \land } \implies A[k] = B[k]$
    
    \item $\I_{\M_1} \equiv \forall \alpha \in ScanReturnSet : (s = \bot \vee pc_s \in \{7, \ldots, 10\})$\\
    $\hphantom{\I_{\M_1} \equiv \forall \alpha \in ScanReturnSet : }\implies \exists (\sigma, f) \in \M : \sigma = \alpha$\\
    $\hphantom{\I_{\M_1} \equiv \forall \alpha \in ScanReturnSet : \implies \exists (\sigma, f) \in \M : \ } \wedge (s \neq \bot \implies f(s) = (\Scan, \bot, \bot)$
    
    \item $\I_{\M_2} \equiv \forall \alpha \in ScanReturnSet : pc_s \in \{11, 12\}$\\
    $\hphantom{\I_{\M_2} \equiv \forall \alpha \in ScanReturnSet : } \implies \exists (\sigma, f) \in M : \sigma = A \wedge f(s) = (\Scan, \bot, \alpha)$
 \end{itemize}
\label{inv:snapshot}
\end{minipage}}\end{figure}

\end{document}